\documentclass[reprint,aps,prl,nofootinbib,twocolumn,superscriptaddress,preprintnumbers]{revtex4-1}
\usepackage{euscript}
\usepackage{amssymb}
\usepackage{amsfonts}
\usepackage{amsbsy}
\usepackage{epsfig}
\usepackage{subfigure}
\usepackage{amsthm}
\usepackage{amscd}
\usepackage{amstext}
\usepackage{verbatim}
\usepackage{amsmath}
\usepackage{url}
\usepackage{bm}
\usepackage[pdftex]{hyperref}
\usepackage{color}

\usepackage{dcolumn}

\newcommand{\be}{\begin{equation}}
\newcommand{\ee}{\end{equation}}
\newcommand{\bea}{\begin{eqnarray}}
\newcommand{\eea}{\end{eqnarray}}
\newcommand{\bega}{\begin{gather}}
\newcommand{\eega}{\end{gather}}

\newcommand{\bi}{\begin{itemize}}
\newcommand{\ei}{\end{itemize}}
\newcommand{\ben}{\begin{enumerate}}
\newcommand{\een}{\end{enumerate}}
\newcommand{\bca}{\begin{cases}}
\newcommand{\eca}{\end{cases}}
\newcommand{\bln}{\begin{align}}
\newcommand{\eln}{\end{align}}
\newcommand{\bst}{\begin{split}}
\newcommand{\est}{\end{split}}
\def\ie{\begin{equation}\begin{aligned}}
\def\fe{\end{aligned}\end{equation}}
\newcommand{\bma}{\le(\begin{matrix}}
\newcommand{\ema}{\end{matrix}\ri)}

\newcommand\ga{{\ensuremath{{\gamma}}}}

\newcommand\de{{\ensuremath{{\delta}}}}

\newcommand\ov{\over}

\def\le{\left}
\def\ri{\right}

\begin{document}

\title{Landau Instability and Soliton Formation}

\author{Shanquan Lan}
\email{lansq@lingnan.edu.cn}
\affiliation{Department of Physics, Lingnan Normal University, Zhanjiang 524048, China}

\author{Hong Liu}
\email{hong$_-$liu@mit.edu}
\affiliation{Center for Theoretical Physics, Massachusetts Institute of Technology, Cambridge, MA 02139, USA}

\author{Yu Tian}
\email{ytian@ucas.ac.cn}
\affiliation{School of Physical Sciences, University of Chinese Academy of Sciences, Beijing 100049, China}
\affiliation{Institute of Theoretical Physics, Chinese Academy of Sciences, Beijing 100190, China}

\author{Hongbao Zhang}
\email{hongbaozhang@bnu.edu.cn}
\affiliation{School of Physics and Astronomy, Beijing Normal University, Beijing 100875, China}
\affiliation{Key Laboratory of Multiscale Spin Physics, Ministry of Education, Beijing Normal University, Beijing 100875, China}
\affiliation{ Theoretische Natuurkunde, Vrije Universiteit Brussel,
 and The International Solvay Institutes, Pleinlaan 2, B-1050 Brussels, Belgium}



\begin{abstract}

Consider at a finite temperature $T$ a superfluid moving with a velocity $v$ relative to the thermal bath or its normal component. From Landau's argument  there exists a critical $v_c (T)$ beyond which excitations can be spontaneously generated and the system becomes unstable. Identifying the final state induced by such an instability has been an outstanding open question. Using holographic duality we perform dynamical simulations of evolutions from initial unstable states, and find that the system settles to a homogenous superfluid state with a final velocity below the critical velocity. The dynamical evolution process appears to be highly chaotic, exhibiting transient turbulence. Nevertheless we are able to identify from the simulations a universal physical mechanism for the reduction of superfluid velocity, in terms of spontaneous nucleation of solitons. We also derive a simple analytic formula which relates the final velocity to the number of solitons nucleated during the evolution.

\end{abstract}

\maketitle


\section{Introduction}

At zero temperature, a superfluid is well known to flow past an obstacle without friction until its velocity $v$ exceeds the Landau critical velocity $v_c$~\cite{Landau,dis2,PS1,PS2}. Beyond $v_c$, excitations can be spontaneously generated, leading to dissipation. 
Such a quantum mechanically macroscopic phenomenon, first discovered in liquid helium \cite{Kapitza,AM, avenel1985, amar1992, burkhart1994, josserand1995}, has been achieved in controllable manner in dilute ultracold atomic gases \cite{raman1999, stiebberger2000, onofrio2000, inouye2001, steinhauer2002, zawitkowski2006, kiyohito2006, miller2007, engels2007, henn2009, neely2010, ramanathan2011, desbuquois2012, baym2012, kwak2023, wright2013, kwon20151, kwon20152, kwon2016, lim2022, delehaye2015, weimer2015}. 
However, if the cross section of the superfluid is large,  excitations at zero temperature are created only near the localized obstacle, while the superfluid bulk can still flow with $v > v_c$ sufficiently far away from the obstacle. In fact, in an infinite space without an obstacle, a superfluid can flow at any velocity without creating excitations due to boost invariance. The situation becomes very different at a finite temperature. For a superfluid moving with a velocity $v$ relative to the ambient thermal bath or its normal component at a finite temperature $T$,  beyond a critical $v_c (T)$ excitations are supposed to be spontaneously generated {\it everywhere}, thus the system will develop a {\it genuine} instability. The nature of the final state resulting from this instability has been a longstanding open question. 

Simulating  the full dynamical evolution of a superfluid at a finite temperature is a challenging problem, for which there is no satisfactory
conventional method. An often-used approach, the dissipative Gross-Pitaevskii equation--where one introduces dissipative effects by hand--is rather crude and requires significant modeling. This motivates us to turn to holographic duality, which converts certain strongly correlated systems of quantum matter to classical gravitational systems in a curved spacetime with one extra spatial dimension\cite{maldacena1999, gubser1998, witten1998}. In this framework, a superfluid at a finite temperature is holographically dual to a hairy black hole, where finite temperature dissipative effects are incorporated from first principles: once the microscopical theory is fixed, all aspects of the superfluid phase are determined without any phenomenological modeling\cite{hartnoll2008, hartnollj2008}. Since its inception, such a holographic model of superfluids has been adopted to address various issues related to superfluid dynamics, including the dynamics associated with the topological defects\cite{keranens2010, keranen2010,  keranen2011, salvio2012,lan2017, xia2019, lan2019, li2020, wittmer2021, ewerz2021, yan2022, lan2023, lan20232, blaise2023, arean2024, an2024, an2024jhep} and quantum turbulence\cite{chesler2013, ewerz2015, du2015, lan2016, wittmer2024, yang2024}. 


In this {\it Letter}, from studying holographic superfluids, not only do we find that a superfluid with an initial velocity $v_i > v_c (T)$ transitions eventually to a homogeneous superfluid state with a final velocity $v_f < v_c (T)$, but also identify the underlying dynamical mechanism, which turns out to be remarkably simple and elegant. 
The dynamical evolution process from the initial unstable supercritical state to the final state appears to be highly chaotic, exhibiting transient turbulence. But the superfluid velocity is reduced by spontaneously nucleating solitons, with the final velocity determined in a simple way from the number of solitons nucleated during the transition process. 
The physical mechanism we unveil is universal, not depending on specific details of a system. It is thus expected to have wide applications in our understanding of a variety of phenomena arising in the non-equilibrium superfluid dynamics.

\section{Holographic setup, dispersion relations and linear instability of superflow}

For our later purpose, we here present the holographic model of superfluids as well as the relevant results regarding the linear instability of superflow, which was firstly obtained in \cite{amado2014}\footnote{Please refer to the Appendix for the details about our numerics, which is different from that used in \cite{amado2014}.}.
As such, we consider a $(2+1)$-dimensional superfluid,
which can be described by
an Abelian-Higgs model~\cite{hartnoll2008,hartnollj2008}
\begin{equation}
\mathcal{L}=-\frac{1}{4}F_{ab}F^{ab}-|D\Psi|^{2}-m^{2}|\Psi|^{2}
\end{equation}
in a $(3+1)$-dimensional AdS black hole spacetime
\begin{equation}
ds^{2}=\frac{L^{2}}{z^{2}}(-f(z)dt^{2}-2dt dz+d\boldsymbol{x}^2) \ .
\end{equation}
Here we are working in the probe limit, where the backreaction of the bulk matter fields onto the geometry is neglected and the corresponding dual normal fluid is frozen. $\boldsymbol{x}=(x,y)$,  $L$ is the AdS radius, and $f(z)=1-(\frac{z}{z_{h}})^{3}$.  $z=z_{h}$ is the black hole  horizon and $z=0$ is the AdS boundary. The AdS black hole has a Hawking temperature $T =\frac{3}{4\pi z_h}$, which gives the temperature of the dual boundary system.  $\Psi$ is a complex scalar field dual to a boundary order parameter $O$, and $D_a=\nabla_a-iA_a$ where $\nabla_a$ is the covariant derivative associated with the metric, and the $U(1)$ gauge field $A_a$  is dual to a conserved global $U(1)$ current $j^a$
under which $O$ is charged. The chemical potential $\mu$ for the global $U(1)$ symmetry is specified by the boundary value of $A_t$, i.e., $\mu=A_t (z=0)$, and the corresponding charge density $\rho$ is given by $\rho=-\partial_zA_t(z=0)$ in the axial gauge $A_z=0$\footnote{More precisely, the chemical potential $\mu=A_t(z=0)$ is valid only for the equilibrium configuration with $A_t$ vanishing at the horizon.}.  Throughout the paper we will keep the total charge of the system fixed.

The system enters a superfluid phase below
some critical temperature $T_c$ when the order parameter $O$ develops a nonzero expectation value, which in the gravity description corresponds to the condensation of $\Psi$. The superfluid dynamics is governed by the equations
\begin{equation}\label{eom}
D_{a}D^{a}\Psi-m^{2}\Psi=0,\quad \nabla_{a}F^{ab}=i(\overline{\Psi}D^{b}\Psi-\Psi\overline{D^{b}\Psi}) \ .
\end{equation}
For definiteness we take $m^2 = -\frac{2}{L^2}$, for which  there are two possible boundary
conditions for $\Psi$, leading to two different types of superfluids with $O$ having dimensions $2$ and $1$, respectively. Below we focus on the one corresponding to
$O$ having dimension $2$. We will denote $\psi = \langle O\rangle$.

Flow of the superfluid component can be generated by turning on a source $\boldsymbol{a}$ for the spatial components of $U(1)$ current $\boldsymbol{j}$. Here and below bold-face letters always denote vectors in boundary spatial directions. With
\begin{gather}
\boldsymbol{\mathcal{J}}= \frac{i}{2}[\psi(\boldsymbol{\partial}+i\boldsymbol{a})\overline{\psi}-\overline{\psi}(\boldsymbol{\partial}-i\boldsymbol{a})\psi] , \nonumber\\
\mathcal{J}_t =\frac{i}{2}[\psi(\partial_t+i\mu) \overline{\psi}-\overline{\psi}(\partial_t -i\mu)\psi] ,
\end{gather}
the superfluid velocity can be written as
\begin{equation}\label{ve}
\boldsymbol{v}=\frac{\boldsymbol{\mathcal{J}}}{{\mathcal J}^
t}=-\frac{\boldsymbol{a}-\boldsymbol{\partial}\theta}{\mu-\partial_t\theta},
\end{equation}
where $\theta$ is the phase of $\psi=|\psi|e^{i\theta}$.
In the gravity description  $\boldsymbol{a}$ can be identified as the boundary value of the bulk gauge field $A_a$ in the $\boldsymbol{x}$ directions.
We will take $\boldsymbol{a}$ to be in the $x$-direction, so is the induced superflow.

\begin{figure}
\begin{center}
\includegraphics[scale=0.5]{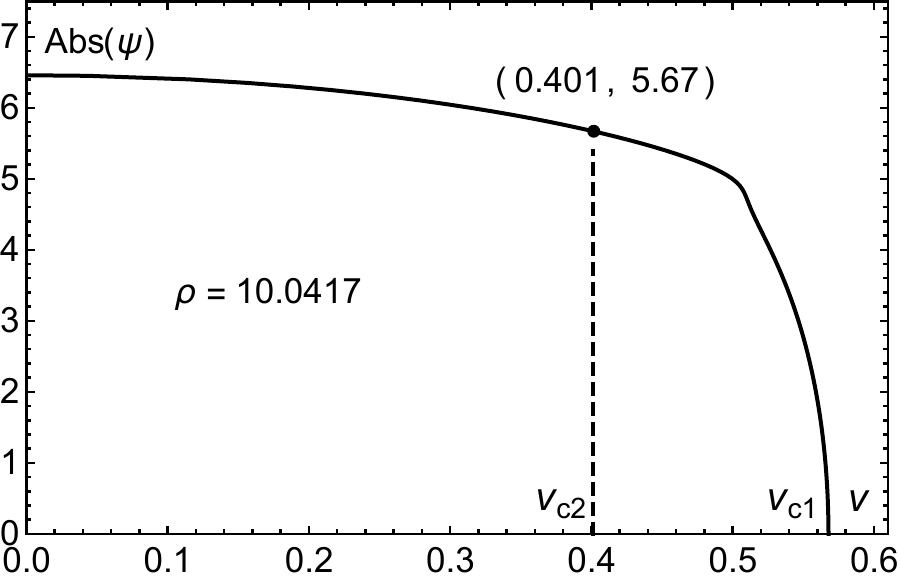}
\end{center}
\caption{The superfluid condensate as a function of the superfluid velocity at $\frac{T}{T_c}=0.637$.  }
\label{condvx}
\end{figure}

With the boundary condition fixed, the superflow solution can be obtained by solving the equations of motion (\ref{eom}).
In FIG.~\ref{condvx}, we show how the magnitude of the superfluid condensate depends on the superfluid velocity.
Notice that the magnitude decreases with the velocity, with the condensate disappearing beyond a certain critical value $v_{c1}$.
The system in fact already becomes unstable at a value $v_{c2} < v_{c1}$, which can be seen from a linear response analysis.
 Due to translational symmetries along the boundary directions, it is convenient to decompose small perturbations around a superflow solution in terms of Fourier modes $ e^{-i \omega t+i\boldsymbol{k}\cdot\boldsymbol{x}}$. Solving linearized equations of motion~(\ref{eom}) we find a discrete spectrum of quasinormal modes $\omega (\boldsymbol{k})$, which are  complex due to dissipations into the thermal bath.  The lowest lying mode is the sound mode and for $v=0$ has the dispersion relation
$\omega (\boldsymbol{k}) =  c_s |\boldsymbol{k}| - i \gamma |\boldsymbol{k}|^2$ (for small $ |\boldsymbol{k}|$),
where $c_s$ is the sound speed and $\gamma$ characterizes its attenuation. For example, for $T/T_c = 0.637$ we have
$c_s =0.63$. With a nonzero $v$, the system is no longer isotropic, accordingly $c_s$ and $\gamma$ become direction-dependent. 
The maximal and minimal values of $c_s$ are achieved in directions  parallel
 (with $k_x > 0, k_y =0$) and anti-parallel (with $k_x < 0, k_y =0$) with the superflow. We will denote them
 respectively as  $c_s^\pm (v)$, and the corresponding values for $\ga$ will be denoted as $\ga^\pm$.

\begin{figure}
\begin{center}
\includegraphics[scale=0.4]{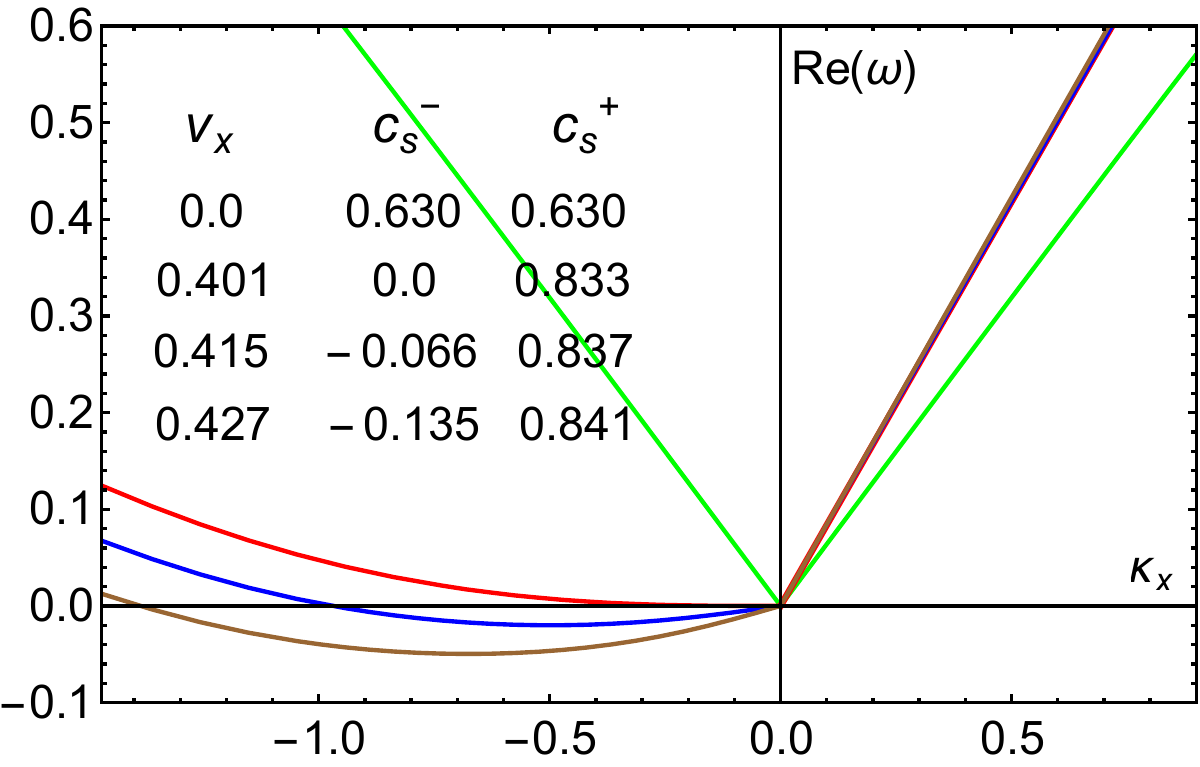}
\includegraphics[scale=0.40]{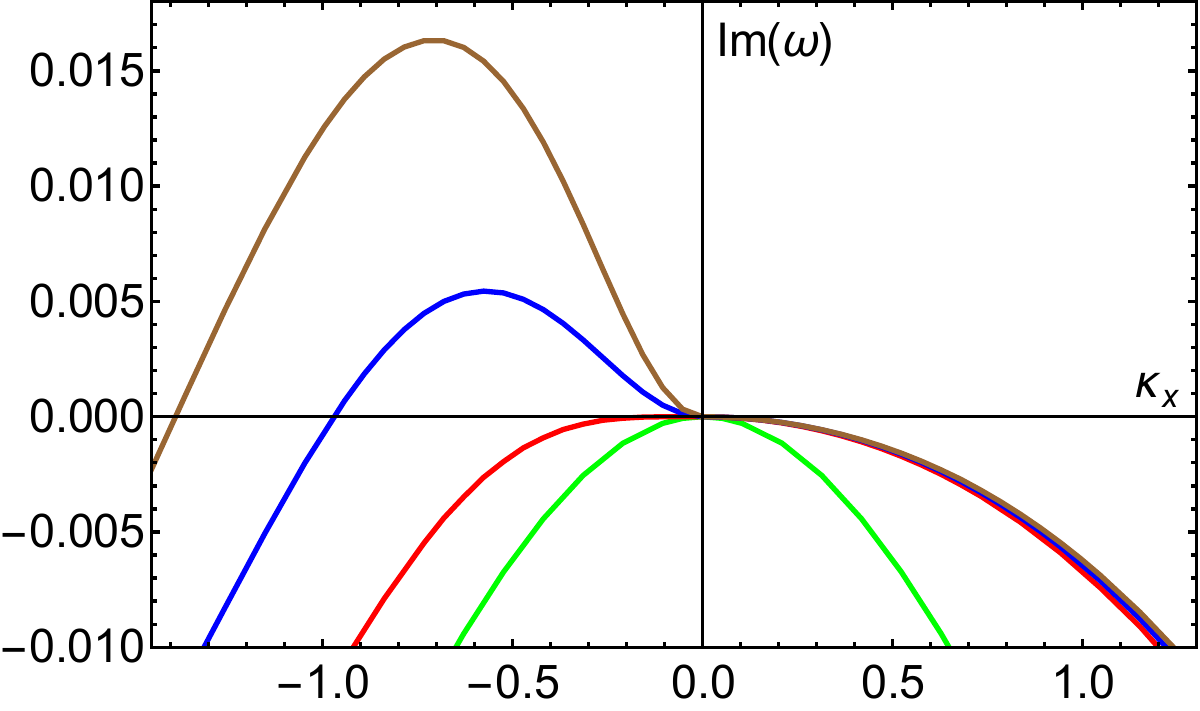}
\end{center}
\caption{The dispersion relation of sound modes for $k_y =0$ at $\frac{T}{T_c}=0.637$, where the green, red, blue, and brown lines are for velocity $v=0, 0.401,0.415, 0.427$. The corresponding values of $c_s^\pm$ are also listed.
The onset of instability is signaled by the red line, where $c_s^- = 0$ and $\gamma^- =0$.
}
\label{qnm}
\end{figure}

\begin{figure}
\begin{center}
\includegraphics[scale=0.60]{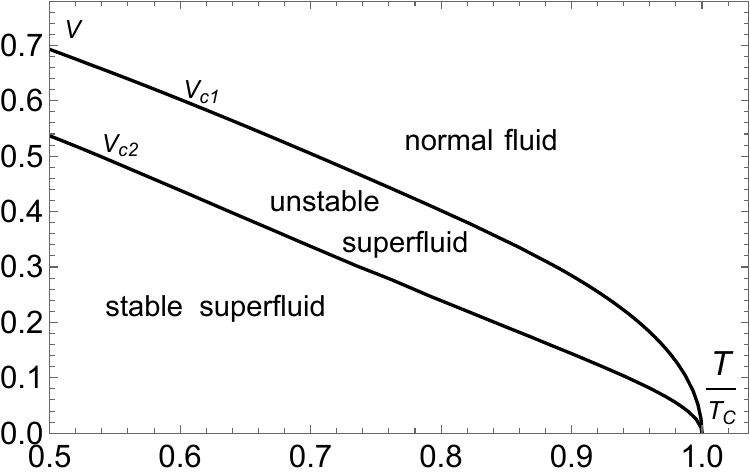}
\end{center}
\caption{The phase diagram for superflow, where the unstable superflow and stable superflow are separated by $v_{c2}$.}
\label{phase}
\end{figure}

In FIG.~\ref{qnm} we plot the dispersion relations with $k_y =0$ for various values of $v$. We notice that as $v$ increases beyond
a certain value $v_{c2} =0.401 $,  $\ga^-$ changes sign and ${\rm Im} \, \omega (\boldsymbol{k})$ becomes positive for sufficiently small $|k_x|$, signaling that the system
becomes unstable. Furthermore, beyond the same value of $v_{c2}$, 
$c_{s}^-$  becomes negative, and thus
the excitation energy becomes negative. This is consistent with the expectation of Laudau's argument. So we also call such an instability as Laudau instability. 
See FIG.~\ref{phase} for the resulting phase diagram of the system.

\section{Full nonlinear simulations, soliton formations and final state of unstable superflow}

We now examine the dynamical process and final state resulting from Laudau instability using full nonlinear simulations.
We use units in which temperature is $T= {3 \ov 4 \pi}$ and work in a $R \times R$
periodic box with $R =30$, which is large enough such that the boundary effect is negligible for the physics we are concerned with here. The system is evolved with an initial configuration of the form
$\Psi_0 (z,\boldsymbol{x}) =\Psi_b (z)e^{i\chi(\boldsymbol{x})}$ where $\Psi_b (z)$ is the background solution for the unstable superflow,  and
\be\label{perb}
\chi(\boldsymbol{x})=c\, \mathrm{Re}\sum_{\boldsymbol{k}}\xi(\boldsymbol{k})e^{i\boldsymbol{k}\cdot\boldsymbol{x}}
\ee
 with $c$ a small real constant and $\xi(\boldsymbol{k})$ a set of $\mathcal{O}(1)$ random complex coefficients.
For comparisons we also consider one-dimensional~(1D) simulations by freezing the dynamics along the $y$ direction.

\begin{figure}
\begin{center}
\includegraphics[scale=0.35]{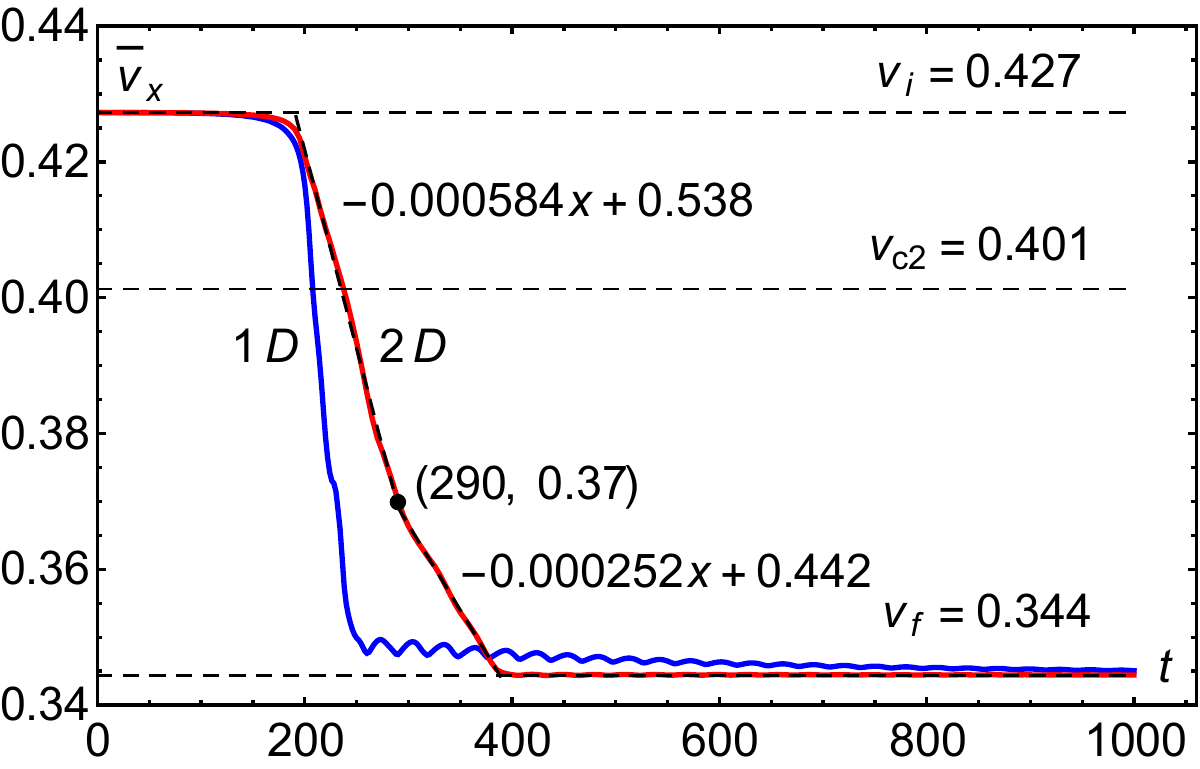}
\end{center}
\caption{The averaged superfluid velocity as a function of time, where the blue and red lines denote the 1D and 2D superflow respectively at $\frac{T}{T_c}=0.637$ with the intermediate horizontal dashed line indicating the corresponding critical velocity. For comparison, we specifically choose simulations such that the 1D and 2D cases have the same final velocity. The velocity decrease in the 2D case can be fitted by two linear segments, with their respective slopes indicated in the plot.
}
\label{velocity}
\end{figure}

In FIG.~\ref{velocity} we plot the time evolution of the average superfluid velocity along the $x$ direction for a typical initial configuration
in both $1$D and $2$D, which shows that after some time, $\bar v_x$ in both cases decreases and settles eventually down to a value less than the critical velocity $v_{c2}$. Notice while the 1D curve exhibits a step-function-like drop,  the 2D curve can be fitted  by two segments of linear decrease. The simulations also show that while the system is highly inhomogeneous and chaotic in the intermediate time, $|\psi|$ becomes homogeneous in the final state.


\begin{figure}
\begin{center}
\includegraphics[scale=0.22]{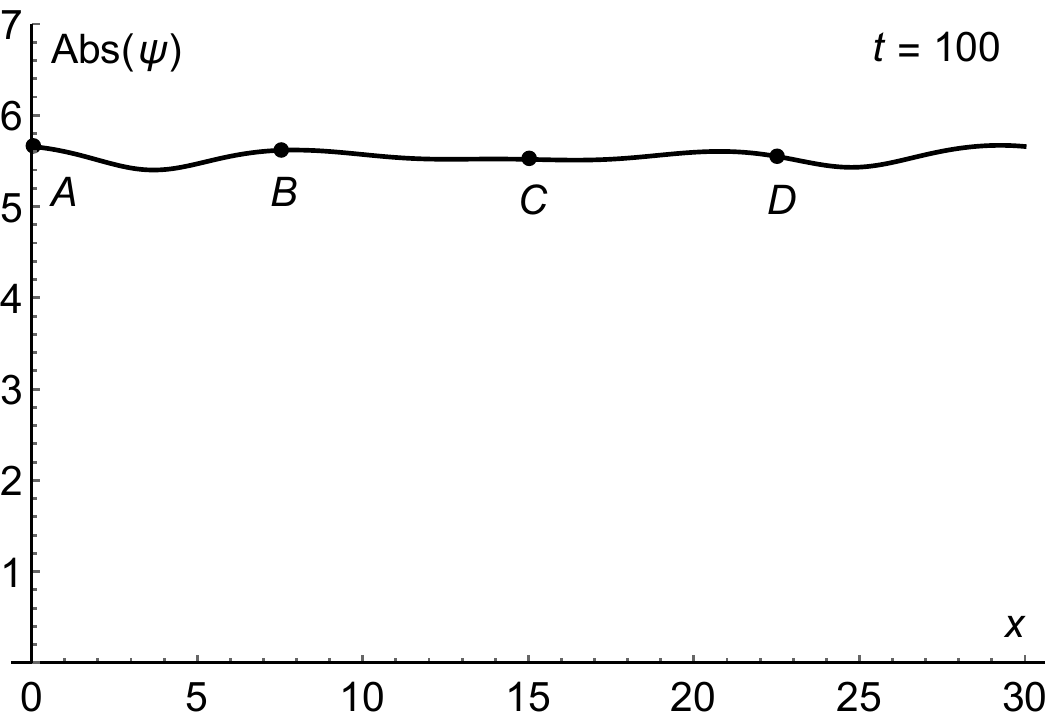}
\includegraphics[scale=0.22]{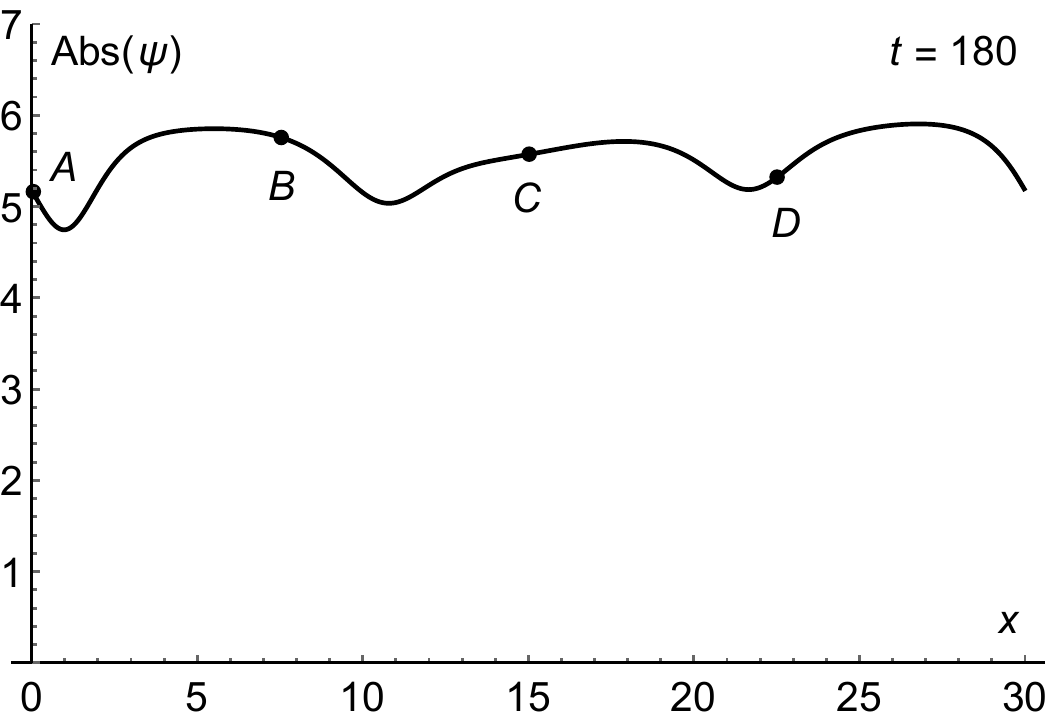}
\includegraphics[scale=0.22]{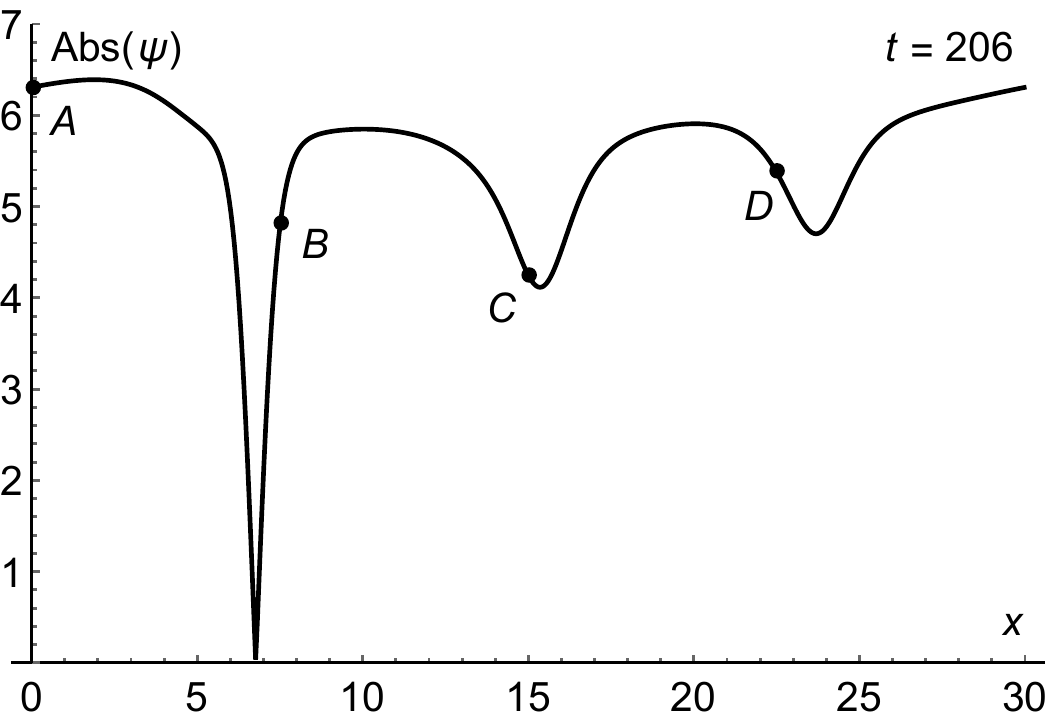}
\includegraphics[scale=0.22]{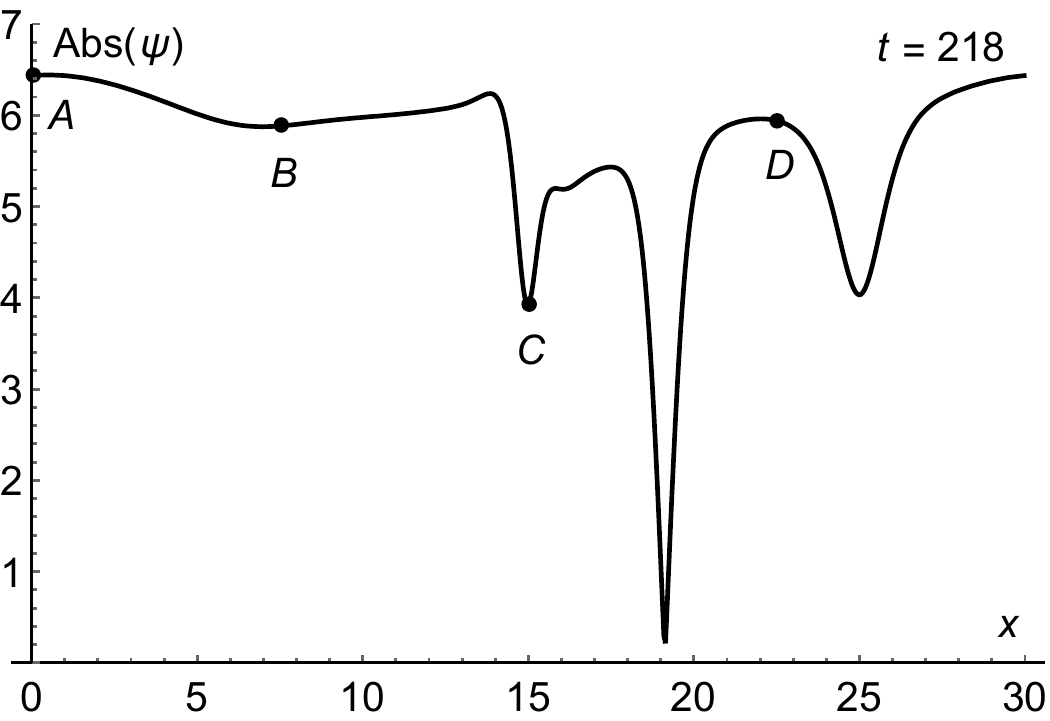}
\includegraphics[scale=0.22]{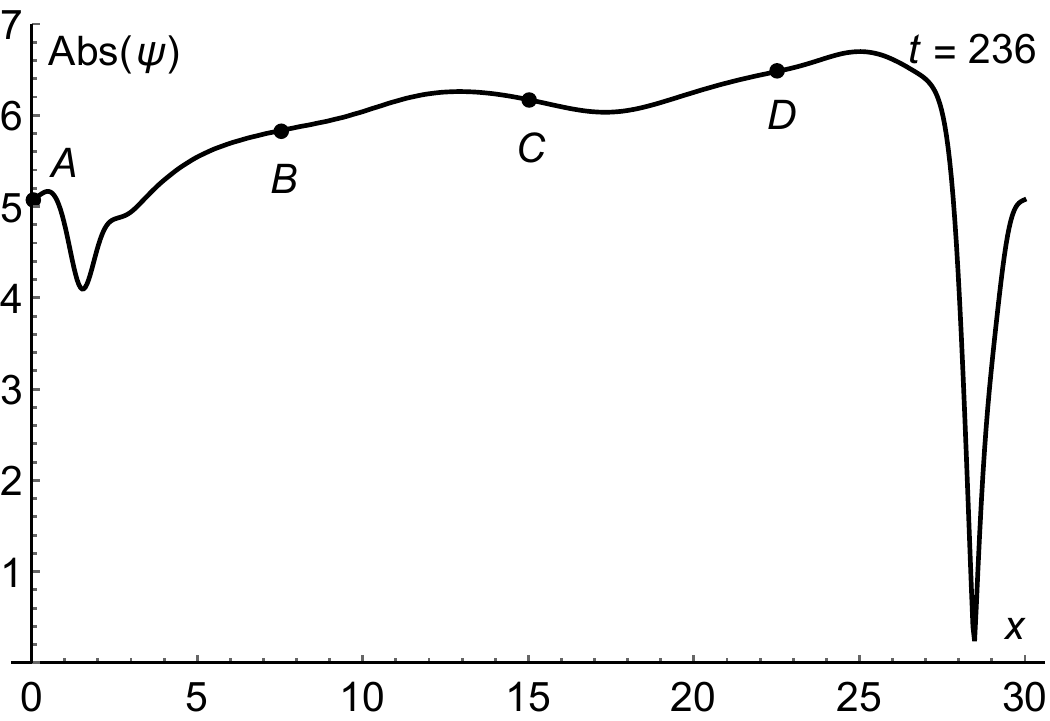}
\includegraphics[scale=0.22]{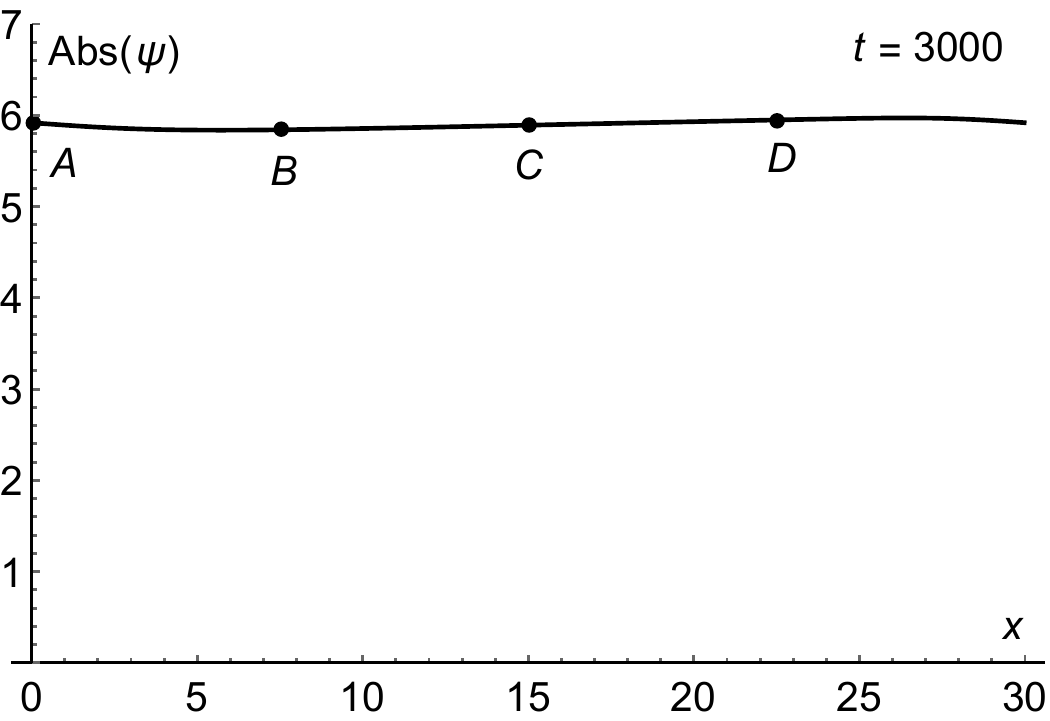}
\end{center}
\caption{Plots of $|\psi (x)|$ at different times for the 1D curve of FIG.~\ref{velocity}. We start with an initial unstable superflow state with small inhomogeneous disturbances.
The inhomogeneity quickly grows as time evolves, but eventually settles to a final state with uniform $|\psi (x)|$.
There is a dark soliton briefly appearing around $t =206, 218, 236$, respectively.
}
\label{fig:psiabs}
\end{figure}

To identify the physical mechanism for the reduction of the superfluid velocity and the physical nature
 of the final state, we will explore the time evolution from two other perspectives. We first consider the 1D case,
 which serves as a simpler example to illustrate the key points. In FIG.~\ref{fig:psiabs} we plot the behavior of $|\psi (x)|$ in the 1D case at various times.
We have specifically included the snapshots at $t =206, 218, 236$, around when a dark soliton (where $|\psi|$ reaches $0$) is formed briefly. Notice the time range of soliton formations precisely coincides with that of the sharp drop in $\bar v_x$ in FIG.~\ref{velocity},
which strongly hints that the physical mechanism for transitioning to the final stable state should have to do with soliton formations.

\begin{figure}
\begin{center}
\includegraphics[scale=0.22]{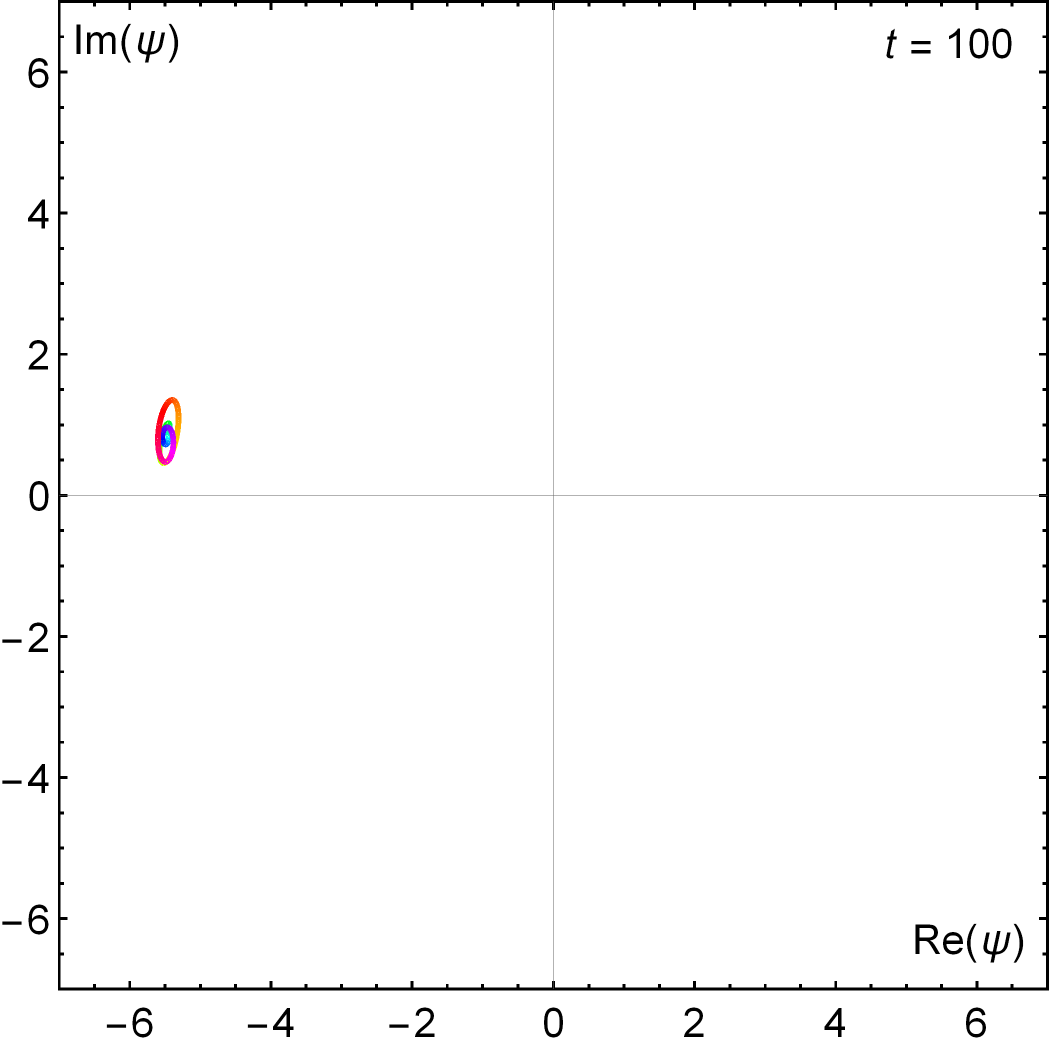}
\includegraphics[scale=0.22]{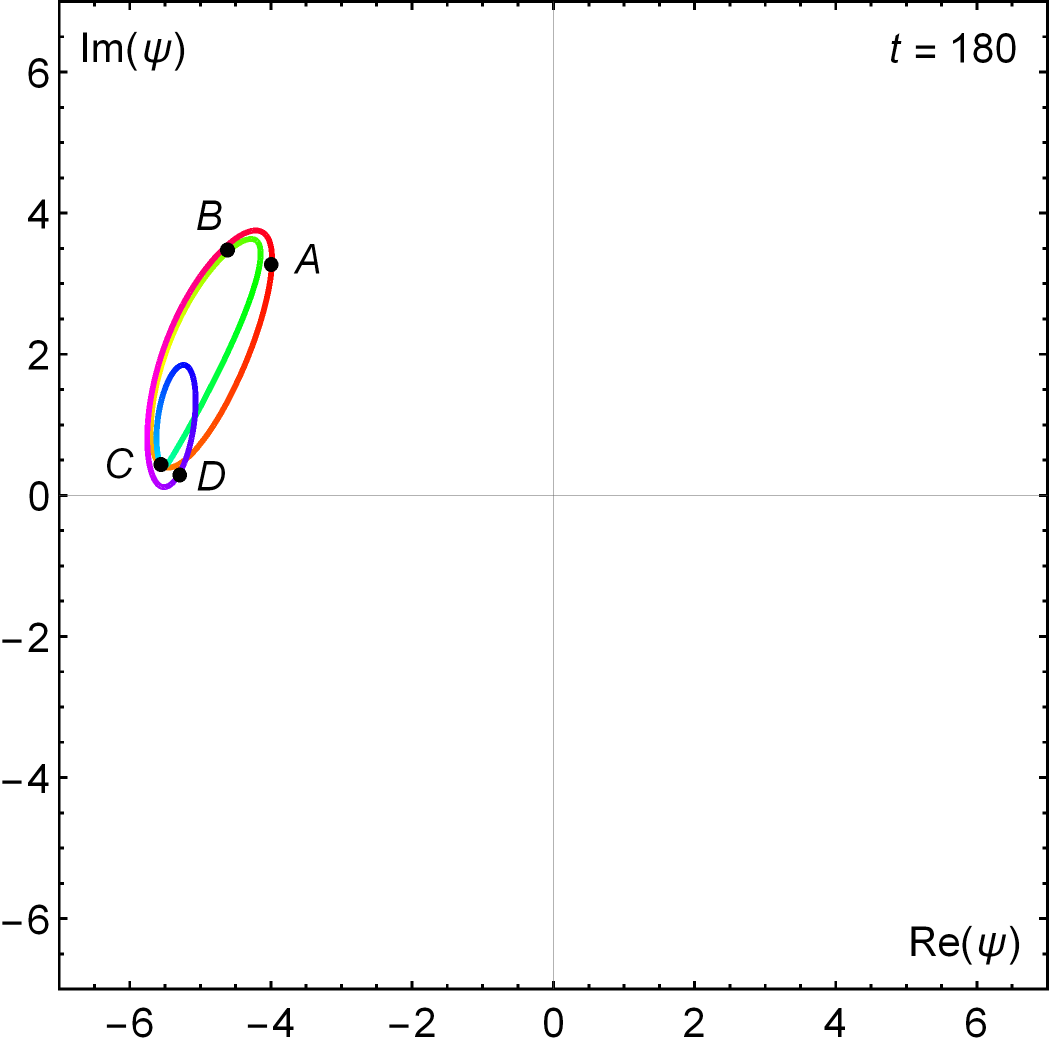}
\includegraphics[scale=0.22]{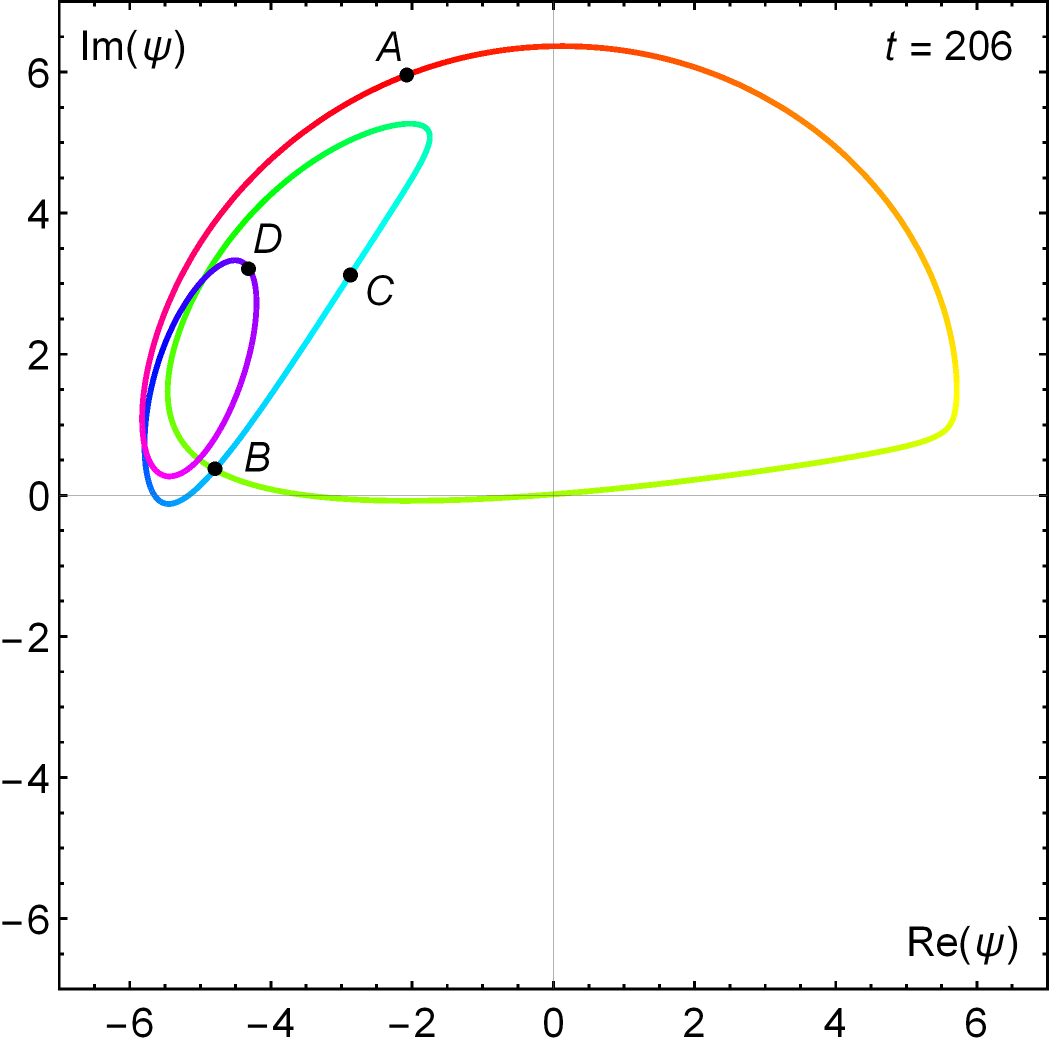}
\includegraphics[scale=0.22]{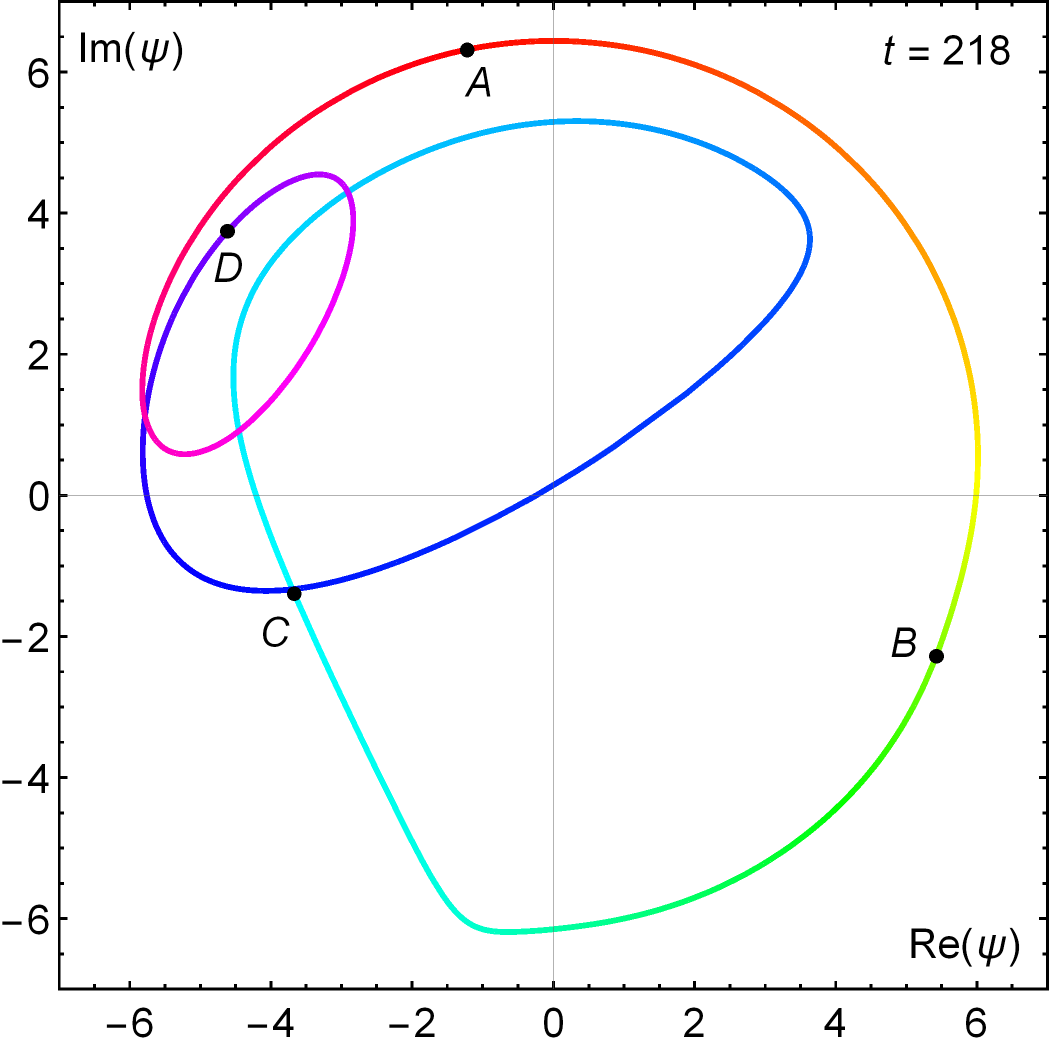}
\includegraphics[scale=0.22]{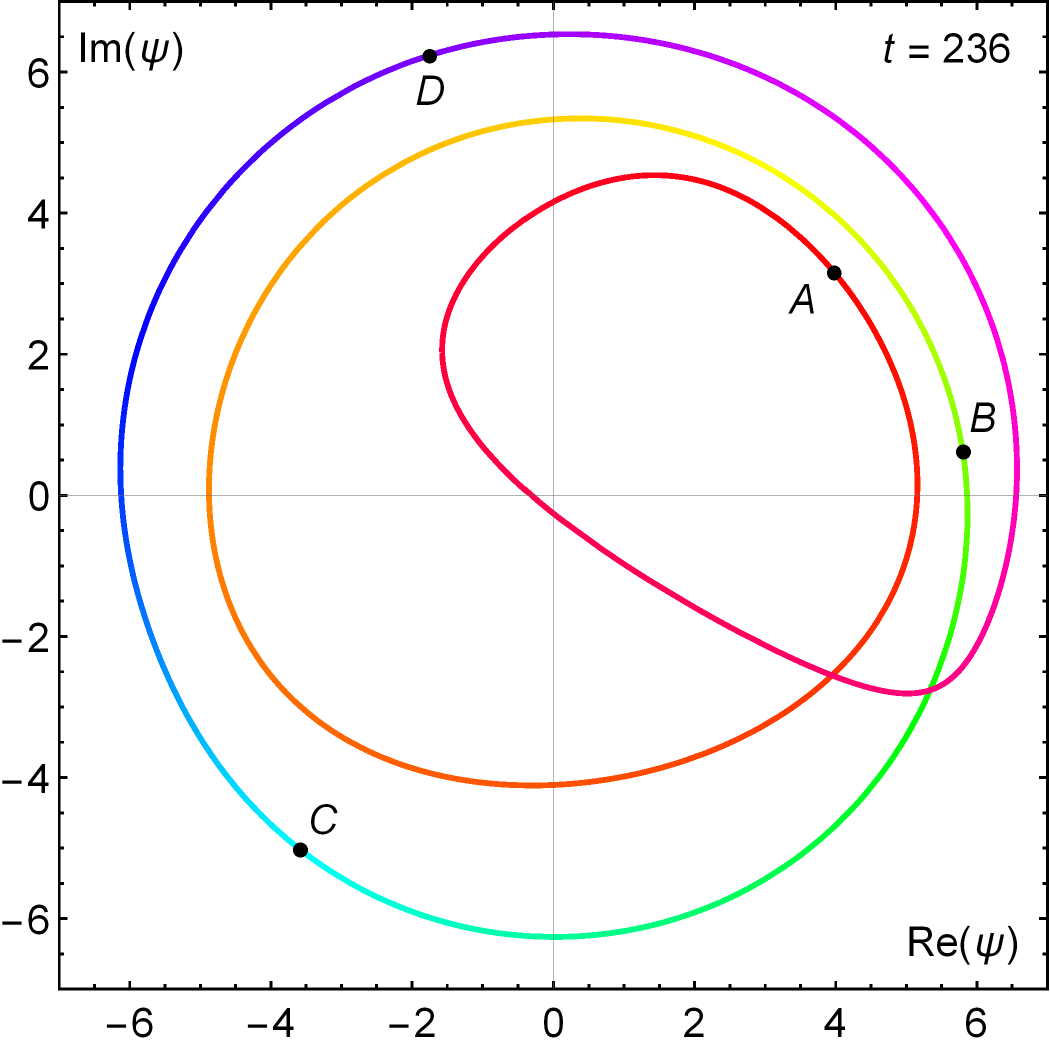}
\includegraphics[scale=0.22]{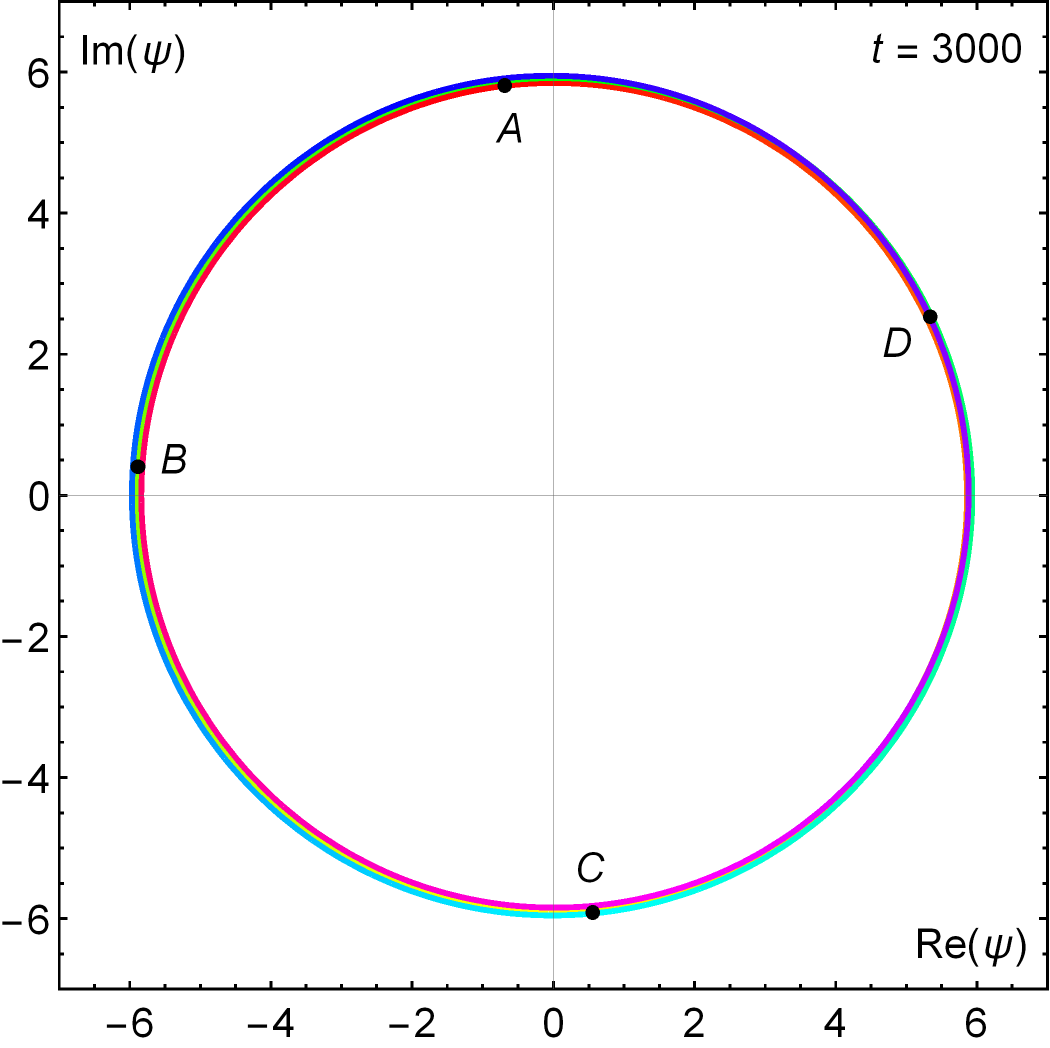}
\end{center}
\caption{The snapshots of superfluid condenstate for the 1D curve of FIG.~\ref{velocity}. Points $A$, $B$, $C$, $D$ denote
the spatial locations $x=0, R/4, R/2, 3R/4$, respectively. The corresponding points are also highlighted in FIG.~\ref{fig:psiabs}. 
}
\label{winding}
\end{figure}

 To sharpen this point,  we plot in FIG.~\ref{winding} the condensate $\psi (t, x)$ on the complex $\psi$-plane, with each plotted point corresponding to the value of $\psi (t, x)$ for some given $t$ and $x$. Since we consider a periodic box,  at a given time all the points  form a closed loop in the complex $\psi$-plane. Such plots have the advantage of showing the variations of both the magnitude and the phase of the condensate over space and time. For example, for a uniform condensate, the whole curve collapses to a single point on the complex $\psi$-plane, whose distance from the origin gives $|\psi|$ and polar angle gives the phase of $\psi$.
In very early times, when the condensate is approximately uniform we see that the whole loop is indeed localized
in a small region of the $\psi$-plane. As the system evolves, the curve quickly expands in a highly irregular manner, reflecting rapid and chaotic growth of inhomogeneity in both the magnitude and phase.

The soliton formations in FIG.~\ref{fig:psiabs} at $t =206, 218, 236$ are reflected in FIG.~\ref{winding} as the loop
 passing through the origin of the complex $\psi$-plane. Now the implication of a soliton formation can be readily understood:
 it  results in a nontrivial winding between the phase $\theta$ and the physical space. More explicitly, before  $t=206$ we have $\theta(R) - \theta (0) =0$, but after that we have $\theta(R) -\theta (0) =- 2 \pi$. Similarly, soliton formations at $t=218$ and $t=236$ generate two additional windings. As indicated from the last plot ($t=3000$) of FIG.~\ref{winding}, both the magnitude $|\psi|$ and the variation of the phase of the final state are expected to be homogeneous along the $x$ direction with $\theta (R) - \theta(0) =-3 \times 2 \pi$.

Now the connection between soliton formation and the drop in velocity is clear. From Eq.~(\ref{ve}), the initial velocity is $v_i =-a_x /\mu_i$, where $\mu_i$ is the initial chemical potential. The final velocity is given by $v_{f} = -{1 \ov \mu_f} (a_x + {2 \pi n \ov R})$ where $n$ is the number of dark solitons formed during the evolution process and $\mu_f$ is the final chemical potential. We thus find a simple elegant relation between the initial and final velocities
\be \label{ehen}
v_{f} ={1 \ov \mu_f} \le(\mu_i v_i - {2 \pi n \ov R}\ri)  \ .
\ee
For the 1D example of FIG.~\ref{velocity}, we have $\mu_i = 6.553$, $\mu_f = 6.306$, and $n=3$. Accordingly, we see that the above equation is well satisfied by 
the resulting $v_i$ and $v_f$ given in FIG.~\ref{velocity} from full nonlinear simulations.



\begin{figure}
\begin{center}
\includegraphics[scale=0.32]{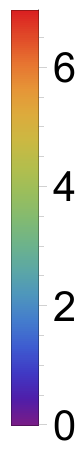}
\includegraphics[scale=0.20]{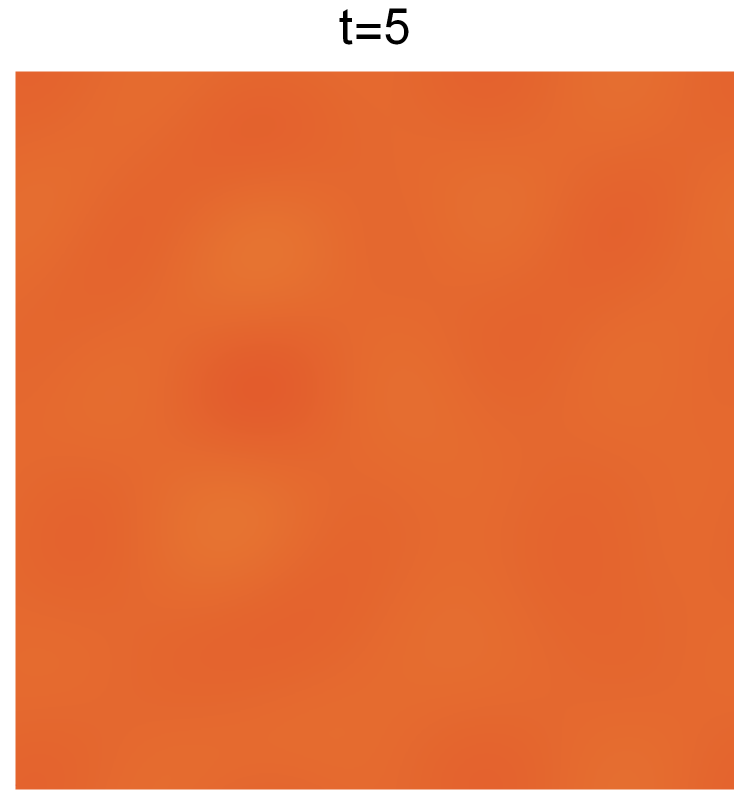}
\includegraphics[scale=0.20]{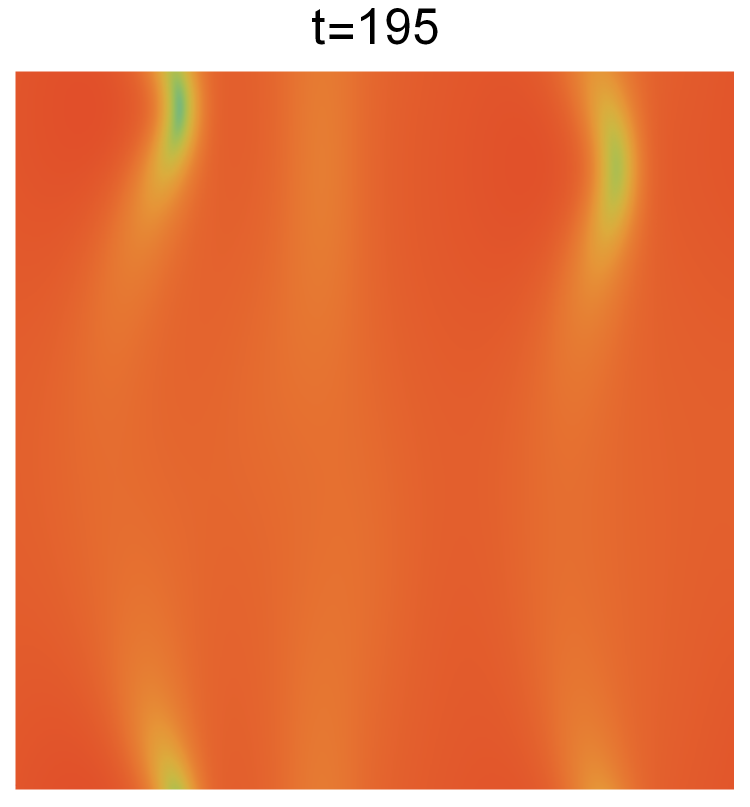}
\includegraphics[scale=0.20]{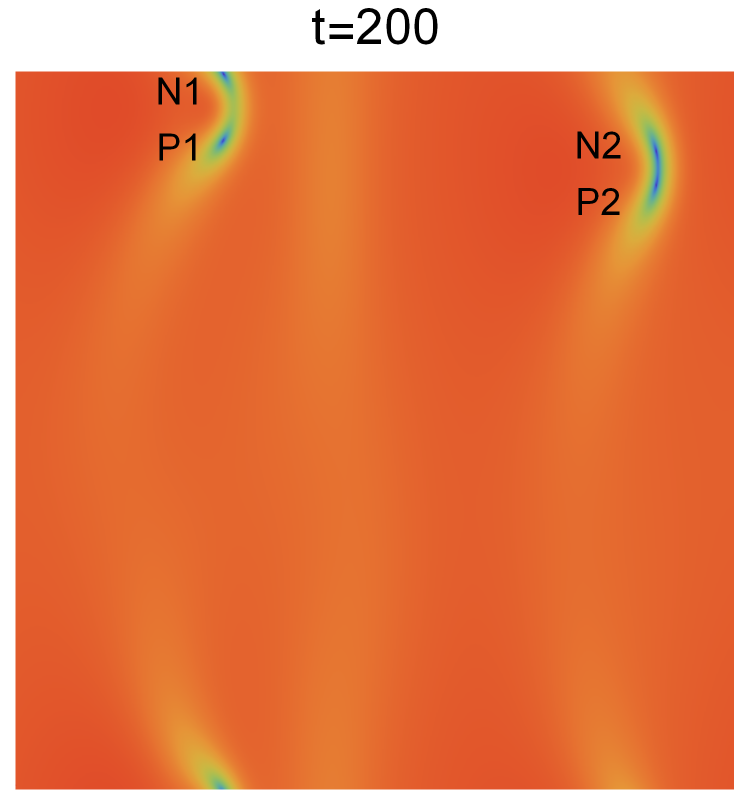}
\includegraphics[scale=0.2]{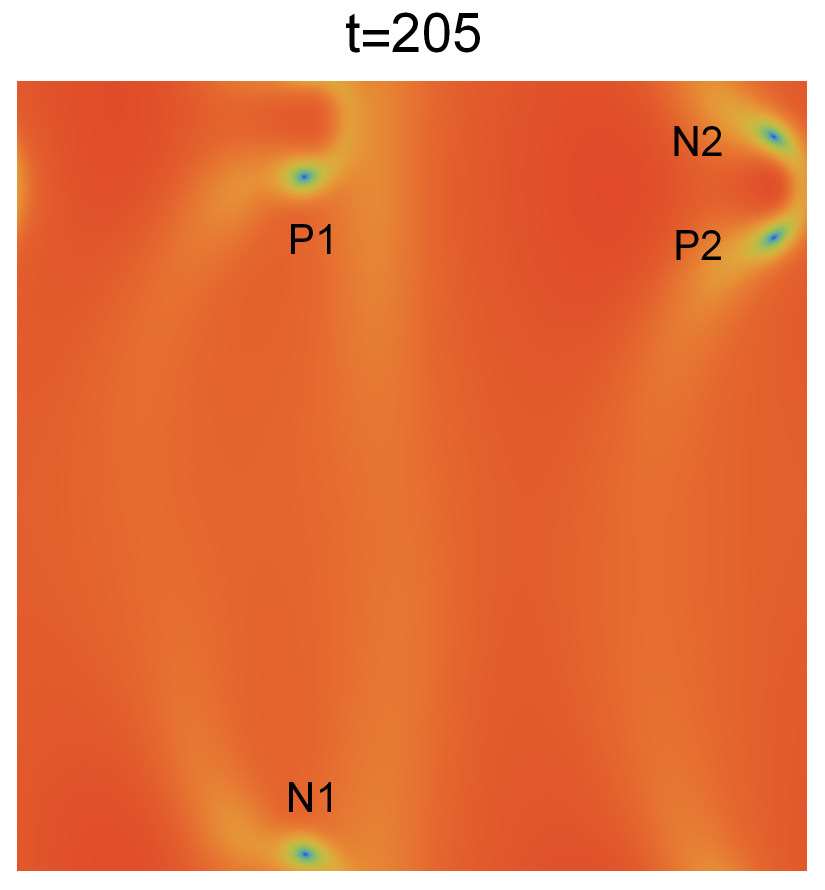}
\includegraphics[scale=0.2]{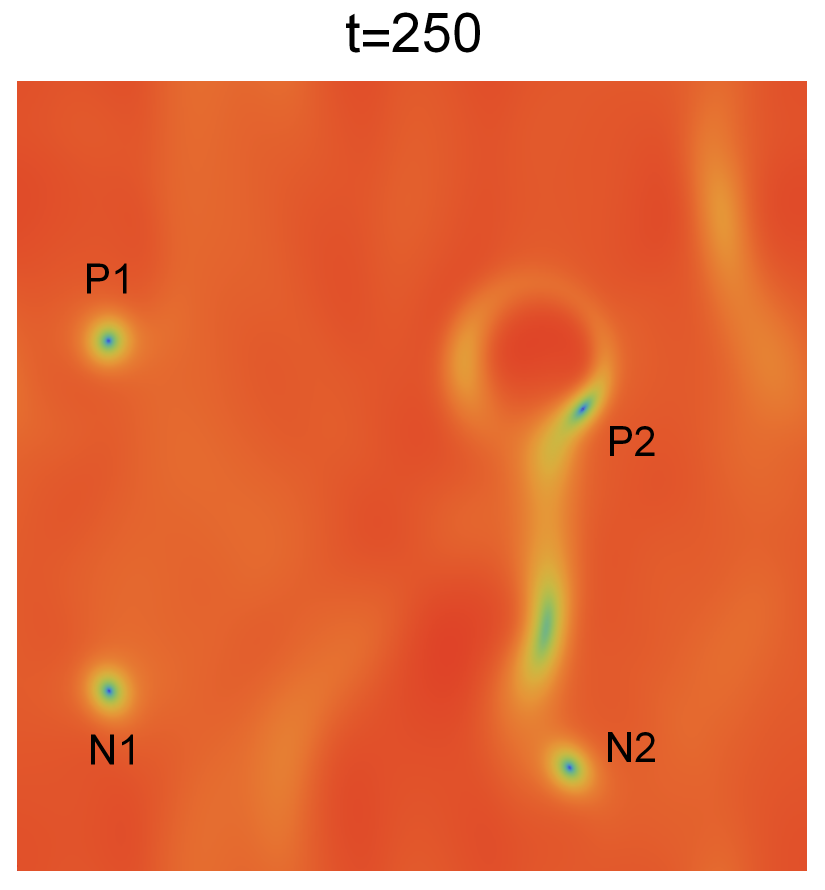}
\includegraphics[scale=0.2]{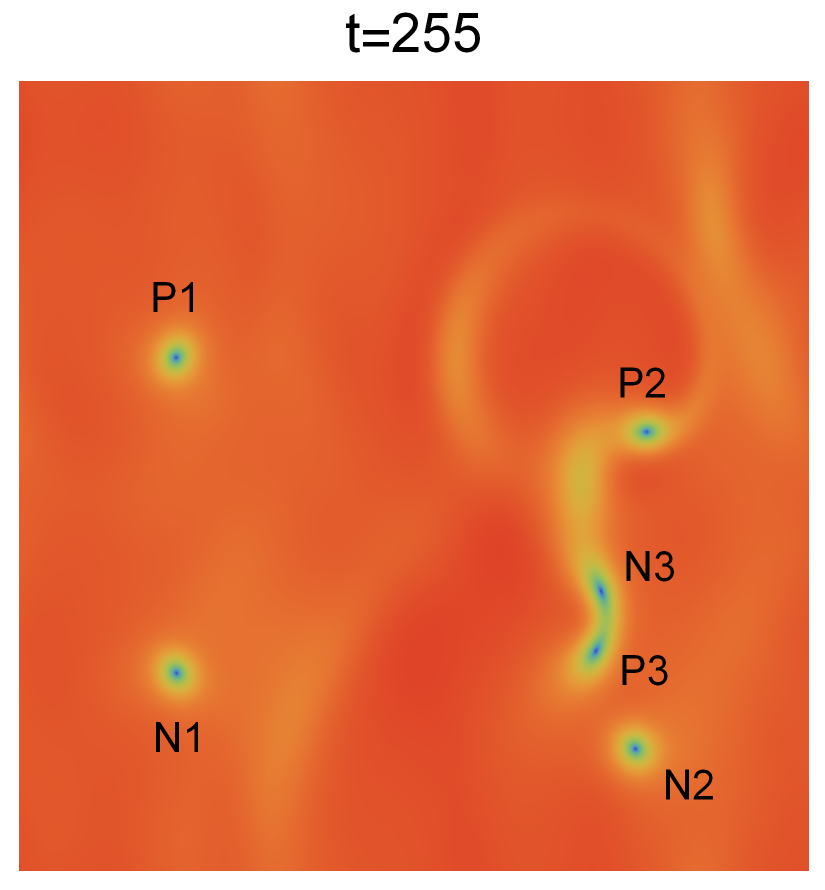}
\includegraphics[scale=0.2]{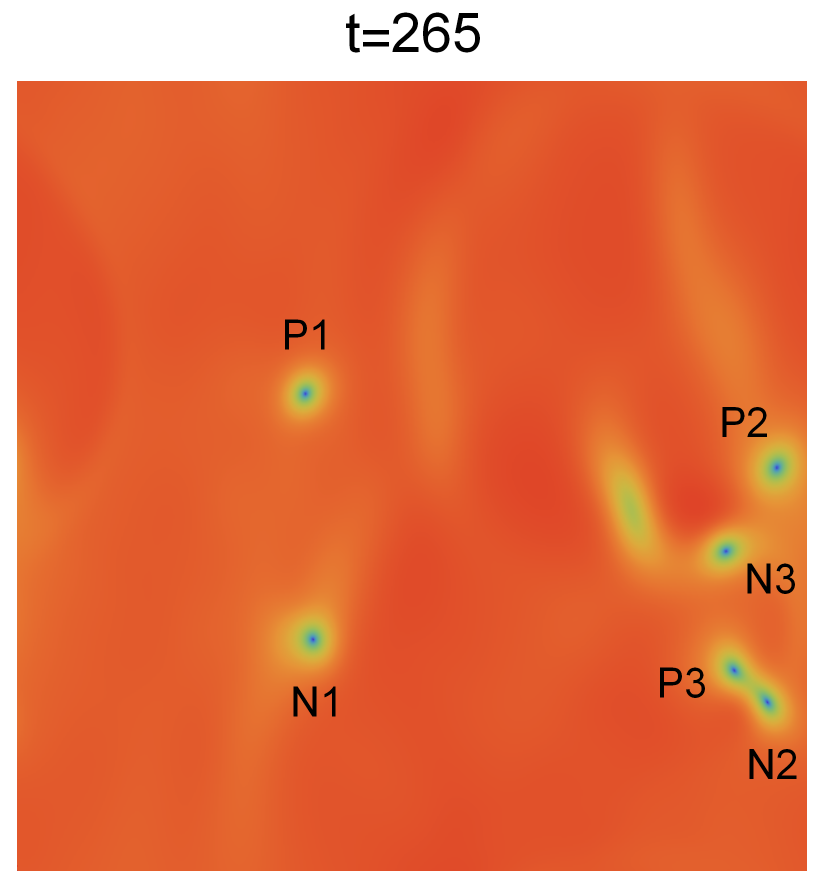}
\includegraphics[scale=0.2]{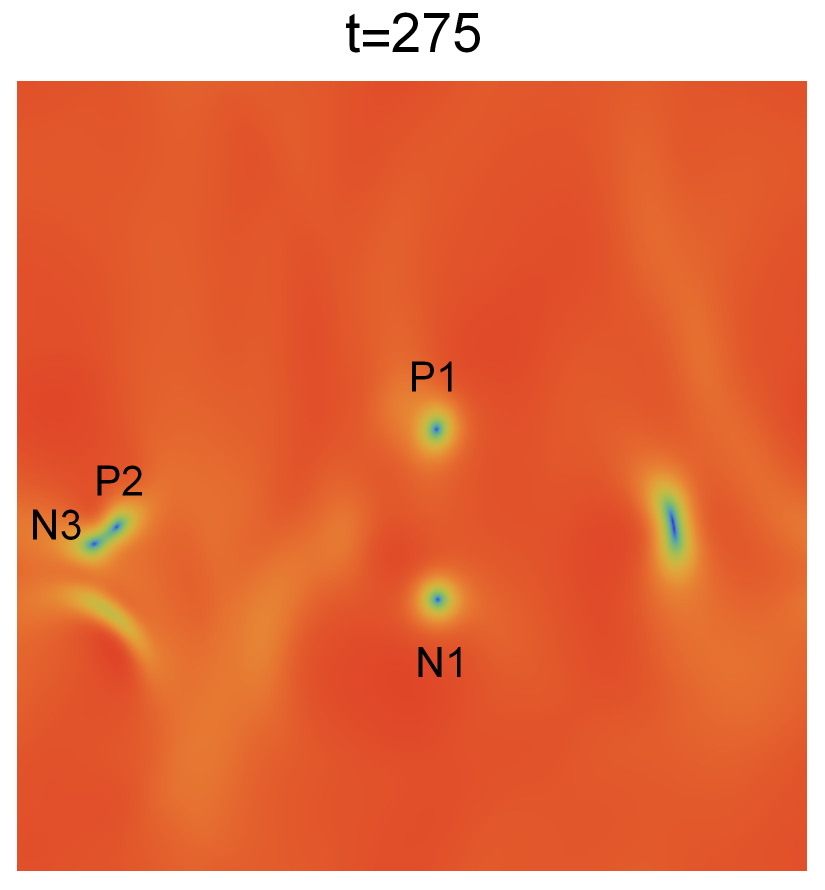}
\includegraphics[scale=0.2]{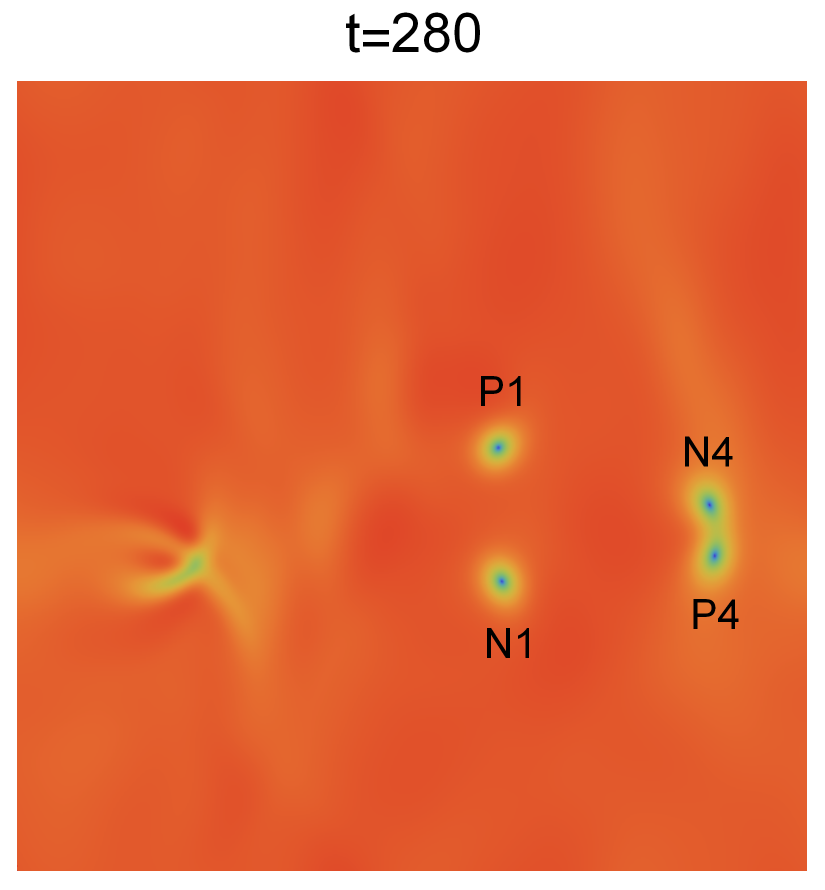}
\includegraphics[scale=0.2]{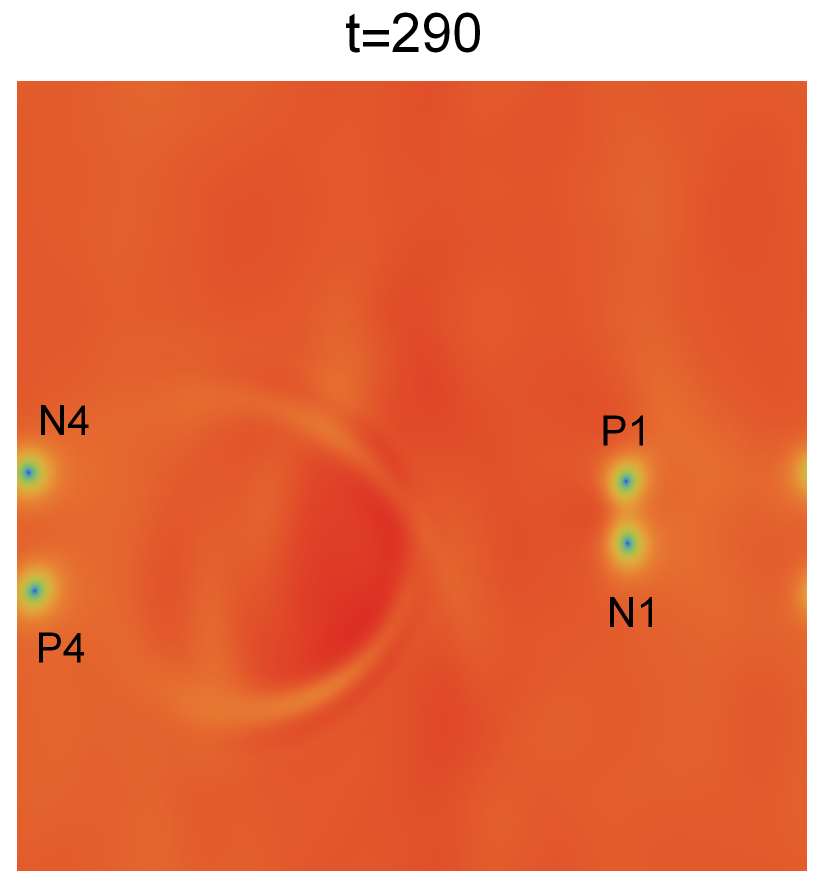}
\includegraphics[scale=0.2]{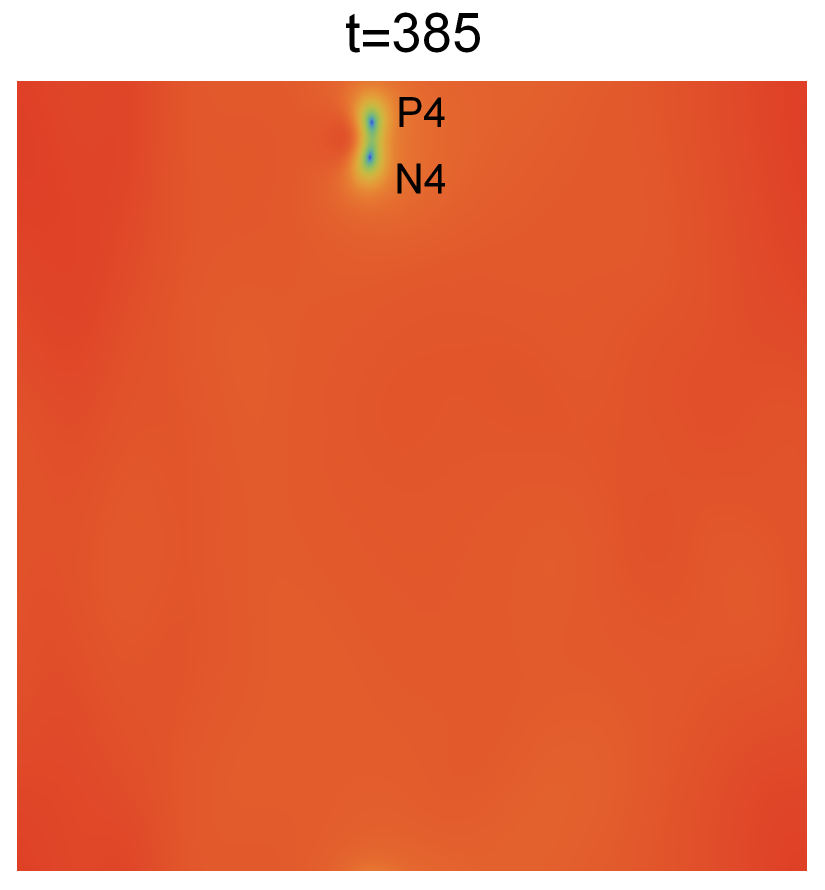}
\includegraphics[scale=0.2]{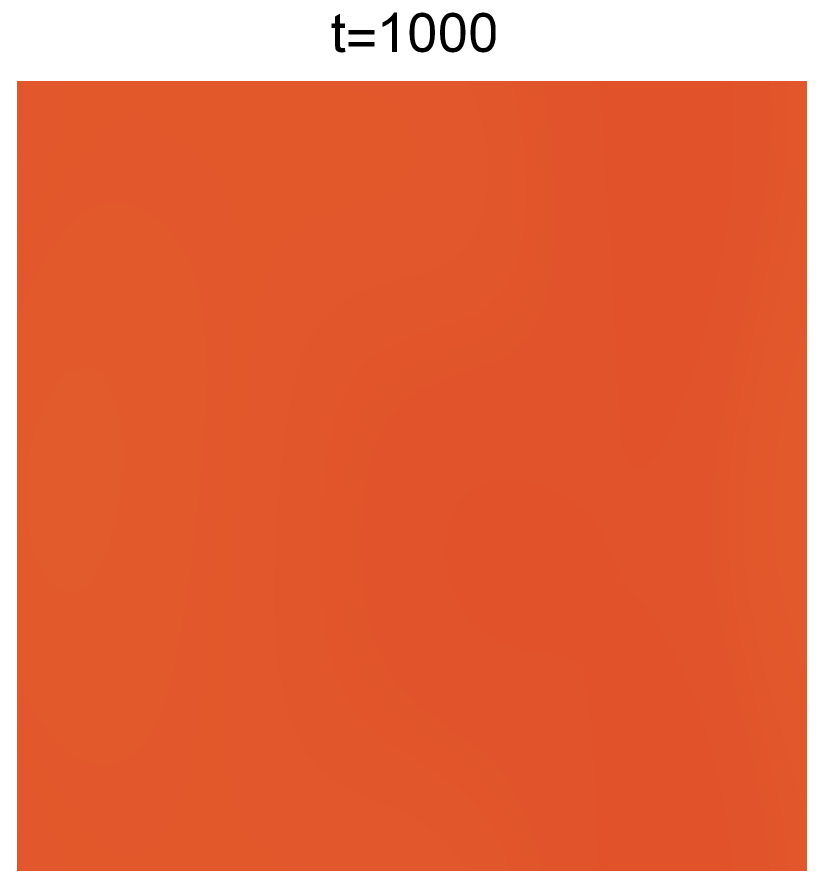}
\end{center}
\caption{The snapshots for the density plot of $2$D superfluid condensate  for the 2D curve of FIG.~\ref{velocity}.
 Vortices and anti-vortices are labeled by $P$ and $N$, respectively.
}
\label{vortex}
\end{figure}

The 2D story works similarly, with formation of dark solitons now replaced by nucleation of vortex-antivortex pairs.
In FIG.~\ref{vortex}, we give the density plot
of the condensate at various times. For example,
two vortex-antivortex pairs are formed around the same time between $t=200$ and $t=205$. When they are formed, the vortex and antivortex move in opposite directions
along the $y$-axis (i.e. perpendicular to the direction of the superflow) with a constant velocity $u_y \approx 0.13$.
This may be understood as a result of opposite Magnus force they each experience, and the balance between the Magnus force and friction. Since the box is periodic, the vortex and antivortex in a pair will meet again and annihilate after $t = {R \ov 2 u_y}$.
They may also annihilate with vortices from other pairs if they happen to meet. All vortices also move along with the superfluid in the $x$-direction.

\begin{figure}
\begin{center}
\includegraphics[scale=0.24]{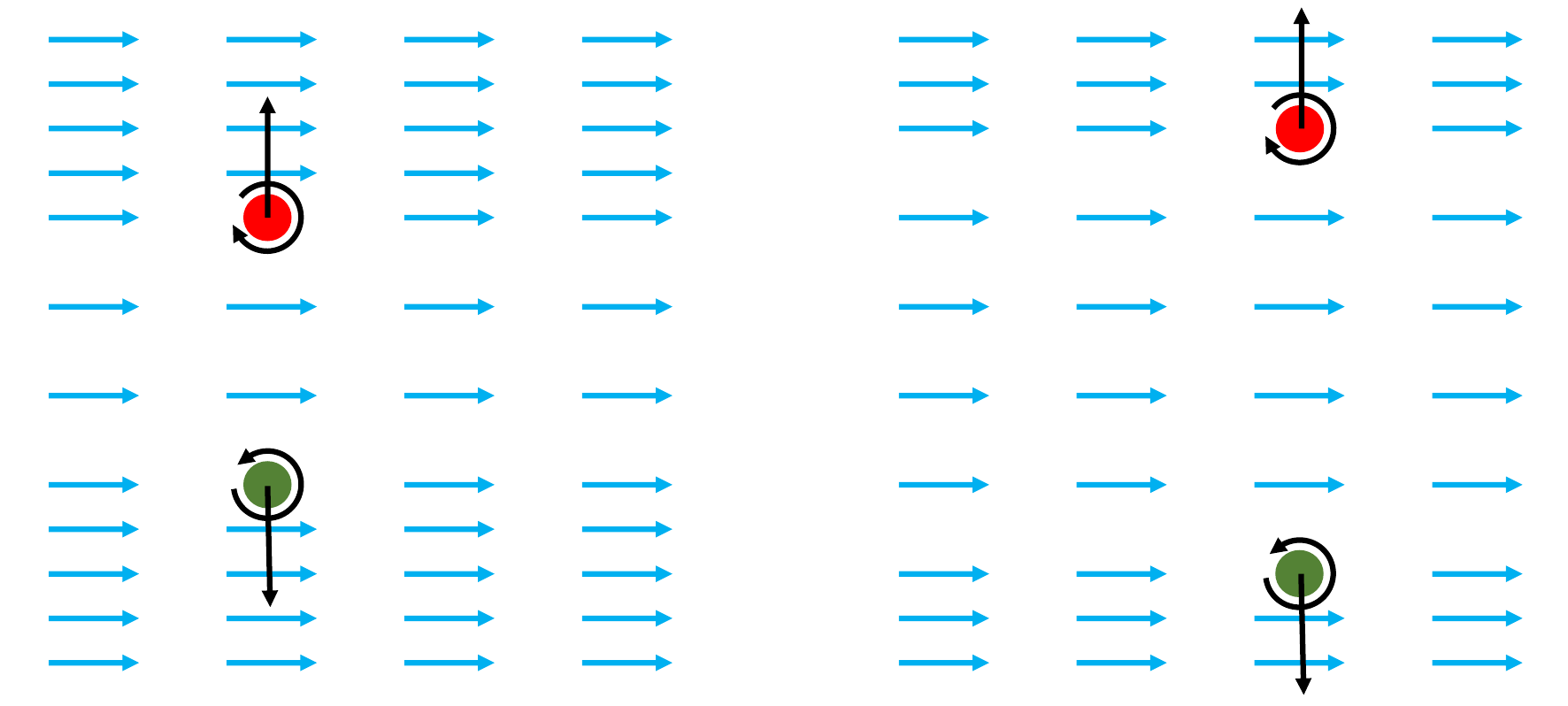} 
\end{center}
\caption{The cartoon for the decrease of the superfluid velocity by the departure of a formed vortex pair from each other.}
\label{cartoon}
\end{figure}

The effect of vortex formations on superfluid flow velocity is as follows: when a vortex (or anti-vortex) passes a horizontal line $y= y_0$, the winding of the condensate in the $x$-direction at $y_0$ reduces by $1$\footnote{ If we make plots like FIG.~\ref{winding} for 2D by restricting to a single value $y=y_0$, then when a vortex passes, the loop passes the origin of the $\psi$-plane and a winding is generated. The sign of the winding change does not depend on whether it is a vortex or anti-vortex because the vortex moves in an opposite direction to the anti-vortex.}.
Thus the superflow velocities in the horizontal strip between the vortex and anti-vortex are reduced, as indicated in FIG.~\ref{cartoon}. When the vortex and anti-vortex meet again and annihilate, the $x$-winding for the whole box will have reduced by $1$.
Suppose that a vortex pair is formed at $t_0$, when the velocity of superflow is $v_0$, and that these are the only vortices in the system,
 then for $0 <\delta t= t-t_0 <  {R \ov 2u_y}$, the average velocity is given by
$\bar v (t) =  {1 \over R} \left(v_0 (R - 2 u_y \de t) + (v_0 - {2 \pi \over \mu R}) 2 u_y \de t \right)
=  v_0 - {4 \pi u_y (t-t_0) \ov \mu R^2}$.
We see that the velocity decreases linearly with $t$ with a slope $- {4 \pi u_y  \ov \mu R^2}$.
The pair annihilates at $t = t_0 + {R \ov 2 u_y}$ after which the average velocity remains constant until the next vortex pair formation.
When there are multiple vortex pairs in the system at a given time, we simply add up their effects.
In general when there are $n$ pairs of vortices in the system, we have
\be
\text{slope} = - {4 \pi n u_y \ov \mu R^2} \ .
\label{heb}
\ee
%
In FIG.~\ref{vortex}, before $t=200$, there is no vortex. During the time period from $t =200$ to $t=290$ there are on average $2$ pairs of vortices at a given time ($3$ pairs appear only briefly). From $t=290$ to $t=400$, there are on average $1$ pair of vortices at a time. After $t=400$ there is no vortex left. We see in FIG.~\ref{velocity}, the 2D plot exhibits indeed two linear regimes in the stated time ranges, with slopes as indicated in the figure. Their values agree reasonably well with the estimates using~\eqref{heb}, which gives
 $-5.6 \times 10^{-4}$ and $-2.8  \times 10^{-4}$ respectively for $n=2$ and $n=1$ (using $\mu \approx 6.5$).

The physical mechanism underlying~\eqref{heb} is similar to that for generation of resistivity in superfluid films and superconductors~\cite{bardeen,halperin} (see also~\cite{Davison:2016hno} for a recent discussion) in that motions of vortices generate local phase shifts in superfluids which degrade the supercurrent.
But among others, there is one important difference. Here a vortex pair only exists for short time intervals, leading to piecewise linear decay rather than exponential decay.

To summarize, in 1D, since solitons are co-dimension one, they lead to sudden drop in the average velocity, while in 2D vortex pair formations lead to piecewise linear decrease. Also notice in FIG.~\ref{velocity} that the 1D curve exhibits a long tail approaching
the final value, which is not present for 2D. The tail can be explained as follows: the solitons in 1D decay through a process similar to self-acceleration~\cite{guo2020}, which eventually leads to sound waves that mostly propagate in the direction of the superflow and it takes a long time for the sound waves to dissipate. We have indeed checked quantitatively that both the decay rate and oscillation frequency of the tail of the 1D curve can be well explained in terms of the dispersion relations of sound waves of the final state. For 2D, since vortices are co-dimension two, the generated sound waves occupy a small fraction of total volume of the system and thus have much less significant effects. 

\section{ Conclusion and discussion}

To conclude, from full nonlinear simulations, not only have we found that an unstable superflow evolves eventually into a homogeneous state with a velocity less than the critical velocity, but also succeeded in identifying generation of winding numbers from dark soliton/vortex formations as the underlying physical mechanism for the velocity reduction. 

In~\cite{caradoc1999, madison2000, aboshaeer2001, hodby2001, aboshaeer2002, penckwitt2002, schweikhard2004}, rotating superfluids were studied and vortex generation was observed near the edge of the superfluid, but not in the interior. This can be well explained by the physical mechanism we have disclosed. Namely, the edge of the rotating superfluid has the largest linear velocity, so the edge has priority over the interior to exceed the critical velocity, whereby Landau instability induced vortices are then formed near the edge to reduce the velocity. The same mechanism also provides a physical explanation for the production of vortices observed in a supersonically expanding ring-shaped Bose-Einstein condensate~\cite{eckel2018}. On the other hand, the resulting nucleation of soliton/vortex from Laudau instability makes the dynamical process exhibit transient turbulent behavior. It is natural to expect that if we keep driving the system so that it remains above the critical velocity, we should find a steady turbulent state. Indeed this is the case and has been observed in a Bose gas \cite{Navon}.

Besides various qualitative implications mentioned above, our findings can in principle be tested quantitatively in necessarily finite temperature atomic systems, albeit ultracold. Furthermore, in light of the implementations of superfluidity in a  room temperature magnon Bose-Einstein condensate and a polariton condensate \cite{Bozhko,nardin2011, grosso2011, amo2011, lerario2017, pigeon2021}, they may be probed in much wider classes of experimental systems.



\section*{Acknowledgements}

We would like to thank Blaise Gout\'eraux, Ted Jacobson, and Tanmay Vachaspati for their informative discussions.
S.L. is supported by Guangdong Basic and Applied Basic Research Foundation of China with Grant No. 2024A1515012552. H.L. is supported by the U.S. Department of Energy, Office of Science, Office of High Energy Physics of U.S. Department of Energy under grant Contract Number DE-SC0012567 and DE-SC0020360 (MIT contract  578218), and by the Packard Foundation award for Quantum Black Holes from Quantum Computation and Holography. Y.T. is partially supported by the National Natural Science Foundation of China with Grants Nos. 12035016, 1237505817 and 12361141825.  H.Z. is partly supported by the National Key Research and Development Program of China with Grant No. 2021YFC2203001 as well as the National Natural Science Foundation of China with Grant Nos. 12075026 and 12361141825.

\section*{Appendix}
Below, we describe the involved numerics in detail for those who are interested.

For convenience in our numerical calculations, we will take $L=1$ as our unit, and fix $z_{h}=1$. Then the relevant results can be obtained by the scaling symmetry of the system. In addition, we define a new function $\Phi=\frac{\Psi}{z}$  and work with the axial gauge $A_{z}=0$, in which the bulk equations of motion can be written explicitly as
\begin{eqnarray}\label{eqphi}
\partial_{t}\partial_{z}\Phi&=&iA_{t}\partial_{z}\Phi+\frac{1}{2}[i\partial_{z}A_{t}\Phi+f\partial^{2}_{z}\Phi+f'\partial_{z}\Phi
\nonumber\\
&&+(\partial-iA)^{2}\Phi-z\Phi],
\end{eqnarray}
\begin{equation}\label{constraint}
\partial_{z}(\partial_{z}A_{t}-\partial\cdot\boldsymbol{A})=i(\overline{\Phi}\partial_{z}\Phi-\Phi\partial_{z}\overline{\Phi}),
\end{equation}
\begin{eqnarray}\label{eqa}
\partial_{t}\partial_{z}\boldsymbol{A}&=&\frac{1}{2}[\partial_{z}(\boldsymbol{\partial} A_{t}+f\partial_{z}\boldsymbol{A})+(\partial^{2}\boldsymbol{A}-\partial\boldsymbol{\partial}\cdot\boldsymbol{A})\nonumber\\
&&-i(\overline{\Phi}\partial\Phi-\Phi\partial\overline{\Phi})]-\boldsymbol{A}\overline{\Phi}\Phi,
\end{eqnarray}
\begin{eqnarray}\label{eqat}
\partial_{t}\partial_{z}A_{t}&&=\partial^{2}A_{t}+f\partial_{z}\boldsymbol{\partial}\cdot\boldsymbol{A}-\partial_{t}\boldsymbol{\partial}\cdot\boldsymbol{A}-2A_{t}\overline{\Phi}\Phi\nonumber\\
&&+if(\overline{\Phi}\partial_{z}\Phi-\Phi\partial_{z}\overline{\Phi})-i(\overline{\Phi}\partial_{t}\Phi-\Phi\partial_{t}\overline{\Phi}).
\end{eqnarray}
As a result,  the asymptotic solution of $A$ and $\Phi$ near the AdS boundary can be expanded as
\begin{equation}\label{asymp}
A_{\mu}=a_{\mu}+b_{\mu}z+o(z),\Phi=\phi+\psi{z}+o(z).
\end{equation}
According to the holographic dictionary, the expectation value of $j$ and $O$ can be obtained explicitly by the variation of renormalized bulk on-shell action with respect to the source as
\begin{eqnarray}\label{current}
\langle j^{\mu}\rangle=\frac{\delta{S_{ren}}}{\delta{a_{\mu}}}=\lim_{z\rightarrow 0}\sqrt{-g}F^{z\mu},
\end{eqnarray}
\begin{eqnarray}\label{operator}
\langle O\rangle=\frac{\delta{S_{ren}}}{\delta{\overline{\phi}}}&=&-\lim_{z\rightarrow 0}z\sqrt{-h}(n_{a}D^a\Psi+\Psi)\nonumber\\
&=&\psi-\dot{\phi}+ia_{t}\phi,
\end{eqnarray}
where the dot denotes the time derivative, and the renormalized action is given by
\begin{equation}
S_{ren}=\int_{\mathcal{M}}\sqrt{-g}\mathcal{L}-\int_{\mathcal{B}}\sqrt{-h}|\Psi|^{2}
\end{equation}
with the counter term added to make the original action finite.

For our purpose,  we set $a_{t}=const, a_{x}=const, a_{y}=0, \phi=0$,  thus Eq.(\ref{eqat}) evaluated at the AdS boundary reduces to
\begin{eqnarray}\label{chargecon}
\partial_{t}\rho=-\partial_{z}\boldsymbol{\partial}\cdot\boldsymbol{A}|_{z=0},
\end{eqnarray}
which is essentially the conservation law of charge current.

With the above boundary conditions and the periodic boundary condition along the $\boldsymbol{x}$ direction,
the full nonlinear simulations are performed by employing the pseudo-spectral method with 28 Chebyshev modes in the $z$ direction and 121 Fourier modes in the $\boldsymbol{x}$ direction, as well as the fourth order Runge-Kutta method in time direction with the time step $\Delta t=0.05$.

On the other hand, the laminar superflow solutions can be obtained by solving the equations of motion with the non-vanishing bulk fields dependent only on $z$, which are simplified as
\begin{eqnarray}
&&f\partial_{z}\theta+A_{t}=0,\\
&&2\partial_{z}\theta\phi^{2}+\partial_{z}^{2} A_{t}=0,\\
&&f\partial^{2}_{z}A_{x}+f'\partial_{z}A_{x}-2A_{x}\phi^{2}=0,\\
&&f\partial^{2}_{z}\phi+f'\partial_{z}\phi-(z+A_{x}^{2}+2A_{t}\partial_{z}\theta+f(\partial_{z}\theta)^{2})\phi=0,\nonumber\\
\end{eqnarray}
where we have rewritten $\Phi$ as $\phi(z) e^{i\theta(z)}$, and specified the $x$ direction as the superflow direction.

The onset of Landau instability of such superflow solutions can be analyzed by the linear response theory. To be more specific,  we first decompose the background complex scalar function into its real and imaginary parts as $\Phi(z)=\phi(z){e}^{i\theta(z)}=\Phi_{r}(z)+i\,\Phi_{i}(z)$, and then write all the perturbation functions in terms of the form $\delta(z)e^{-i\omega t+ik cos\alpha x+ik sin\alpha y}$. As a result, the linearized perturbation equations can be expressed as
\begin{eqnarray}
0&=&(z+k^{2}+A_{x}^{2}+(3z^{2}-2i\omega)\partial_{z}-f\partial_{z}^{2})\delta\Phi_{r}\nonumber\\
&&+(\partial_{z}A_{t}-2i k\cos\alpha A_{x}+2A_{t}\partial_{z})\delta\Phi_{i}\nonumber\\
&&+(2\partial_{z}\Phi_{i}+\Phi_{i}\partial_{z})\delta A_{t}+(2A_{x}\Phi_{r}-ik\cos\alpha\Phi_{i})\delta A_{x}\nonumber\\
&&-ik\sin\alpha\Phi_{i}\delta A_{y},
\end{eqnarray}
\begin{eqnarray}
0&=&(-\partial_{z}A_{t}+2i k \cos\alpha A_{x}-2A_{t}\partial_{z})\delta\Phi_{r}\nonumber\\
&&+(z+k^{2}+A_{x}^{2}+(3z^{2}-2i\omega)\partial_{z}-f\partial_{z}^{2})\delta\Phi_{i}\nonumber\\
&&+(-2\partial_{z}\Phi_{r}-\Phi_{r}\partial_{z})\delta A_{t}+(2A_{x}\Phi_{i}+ik\cos\alpha\Phi_{r})\delta A_{x}\nonumber\\
&&+ik\sin\alpha\Phi_{r}\delta A_{y},
\end{eqnarray}
\begin{eqnarray}
0&=&(-2i\omega\Phi_{i}+4A_{t}\Phi_{r})\delta\Phi_{r}+(2i\omega\Phi_{r}+4A_{t}\Phi_{i})\delta\Phi_{i}\nonumber\\
&&+(k^{2}+2(\Phi_{r}^{2}+\Phi_{i}^{2})-i\omega\partial_{z}-f\partial_{z}^{2})\delta A_{t}+\omega k\sin\alpha\delta A_{x}\nonumber\\
&&+\omega k\cos\alpha\delta A_{y},
\end{eqnarray}
\begin{eqnarray}
0&=&(4A_{x}\Phi_{r}+2ik\sin\alpha\Phi_{i})\delta\Phi_{r}+(4A_{x}\Phi_{i}-2ik\sin\alpha\Phi_{r})\delta\Phi_{i}\nonumber\\
&&-ik\sin\alpha\partial_{z}\delta A_{t}
+(2(\Phi_{r}^{2}+\Phi_{i}^{2})+(k\cos\alpha)^{2}\nonumber\\
&&+(3z^{2}-2i\omega)\partial_{z}-f\partial_{z}^{2})\delta A_{x}-k^{2}sin\alpha\cos\alpha\delta A_{y},
\end{eqnarray}
\begin{eqnarray}
0&=&2ik\cos\alpha\Phi_{i}\delta\Phi_{r}-2ik\cos\alpha\Phi_{r}\delta\Phi_{i}-ik\cos\alpha\partial_{z}\delta A_{t}\nonumber\\
&&-k^{2}\sin\alpha\cos\alpha\delta A_{x}
+(2(\Phi_{r}^{2}+\Phi_{i}^{2})+(k\sin\alpha)^{2}\nonumber\\
&&+(3z^{2}-2i\omega)\partial_{z}-f\partial_{z}^{2})\delta A_{y}.
\end{eqnarray}
The corresponding quasinormal modes are extracted by solving the above generalized eigenvalue problem on top of the background superflow solution.

\end{document}